\documentclass[sigconf]{acmart}

\AtBeginDocument{%
  \providecommand\BibTeX{{%
    \normalfont B\kern-0.5em{\scshape i\kern-0.25em b}\kern-0.8em\TeX}}}

\copyrightyear{2025}
\acmYear{2025}
\setcopyright{cc}
\setcctype{by}
\acmConference[CHI '25]{CHI Conference on Human Factors in Computing Systems}{April 26-May 1, 2025}{Yokohama, Japan}
\acmBooktitle{CHI Conference on Human Factors in Computing Systems (CHI '25), April 26-May 1, 2025, Yokohama, Japan}\acmDOI{10.1145/3706598.3713760}
\acmISBN{979-8-4007-1394-1/25/04}

\usepackage{multirow}
\usepackage{soul}
\usepackage[utf8]{inputenc}
\usepackage[linesnumbered,ruled,vlined]{algorithm2e}

\begin{document}


\title{Proactive Conversational Agents with Inner Thoughts}



\author{Xingyu Bruce Liu}
\orcid{0000-0002-6988-5471} 
\affiliation{%
  \institution{University of California, Los Angeles}
  \city{Los Angeles}
  \state{CA}
  \country{USA}}
 \email{xingyuliu@ucla.edu}

\author{Shitao Fang}
\affiliation{%
  \institution{The University of Tokyo}
  \city{Tokyo}
  \country{Japan}}
 \email{fst@iis-lab.org}

 \author{Weiyan Shi}
\affiliation{%
  \institution{Northeastern University}
  \city{Boston}
  \state{MA}
  \country{USA}}
 \email{we.shi@northeastern.edu}

\author{Chien-Sheng Wu}
\affiliation{%
  \institution{Salesforce AI}
  \city{Palo Alto}
  \state{CA}
  \country{USA}}
 \email{wu.jason@salesforce.com}
 
\author{Takeo Igarashi}
\affiliation{%
  \institution{The University of Tokyo}
  \city{Tokyo}
  \country{Japan}}
 \email{takeo@acm.org}

\author{Xiang `Anthony' Chen}
\affiliation{%
  \institution{University of California, Los Angeles}
  \city{Los Angeles}
  \state{California}
  \country{USA}}
\email{xac@ucla.edu}


\definecolor{darkgreen}{rgb}{0,0.5,0}
\definecolor{orange}{rgb}{1,0.5,0}

\definecolor{teal}{rgb}{0,0.5,0.5}
\definecolor{darkpurple}{rgb}{0.5, 0, 0.5}
\definecolor{burntorange}{rgb}{0.8, 0.3, 0}
\definecolor{forestgreen}{rgb}{0.13, 0.55, 0.13}
\definecolor{goldenrod}{rgb}{0.85, 0.65, 0.13}

\definecolor{level-1}{RGB}{215,25,28}
\definecolor{level-2}{RGB}{241, 90, 41}
\definecolor{level-3}{RGB}{251, 196, 64}
\definecolor{level-4}{RGB}{166,217,106}
\definecolor{level-5}{RGB}{26,150,65}


\newcommand{\rrabox}{\colorbox{darkpurple}{\textcolor{white}{\#1}} \textcolor{darkpurple}}
\newcommand{\rra}[1]{\textcolor{darkpurple}{\rrabox #1}}
\newcommand{\rrbbox}{\colorbox{goldenrod}{\textcolor{white}{\#2}} \textcolor{goldenrod}}
\newcommand{\rrb}[1]{\textcolor{goldenrod}{\rrbbox #1}}
\newcommand{\rrcbox}{\colorbox{teal}{\textcolor{white}{\#3}} \textcolor{teal}}
\newcommand{\rrc}[1]{\textcolor{teal}{\rrcbox #1}}
\newcommand{\rrdbox}{\colorbox{burntorange}{\textcolor{white}{\#4}} \textcolor{burntorange}}
\newcommand{\rrd}[1]{\textcolor{burntorange}{\rrdbox #1}}
\newcommand{\rr}[1]{\textcolor{forestgreen}{#1}}

\renewcommand{\rrabox}[1]{#1}
\renewcommand{\rra}[1]{#1}
\renewcommand{\rrbbox}[1]{#1}
\renewcommand{\rrb}[1]{#1}
\renewcommand{\rrcbox}[1]{#1}
\renewcommand{\rrc}[1]{#1}
\renewcommand{\rrdbox}[1]{#1}
\renewcommand{\rrd}[1]{#1}
\renewcommand{\rr}[1]{#1}

\newcommand{\eg}{\textit{e.g., }}
\newcommand{\ie}{\textit{i.e., }}
\newcommand{\etal}{et al. }

\newcommand {\systemname}{}

\begin{abstract}
One of the long-standing aspirations in conversational AI is to allow them to autonomously take initiatives in conversations, \ie being \textit{proactive}. This is especially challenging for multi-party conversations.  
Prior NLP research focused mainly on predicting the next speaker from contexts like preceding conversations. 
In this paper, we demonstrate the limitations of such methods and rethink what it means for AI to be proactive in multi-party, human-AI conversations.
We propose that just like humans, rather than merely reacting to turn-taking cues, a proactive AI formulates its own \textit{inner thoughts} during a conversation, and seeks the right moment to contribute. 
Through a formative study with 24 participants and inspiration from linguistics and cognitive psychology, we introduce the Inner Thoughts framework. 
Our framework equips AI with a continuous, covert train of thoughts in parallel to the overt communication process, which enables it to proactively engage by modeling its \textit{intrinsic motivation} to express these thoughts. 
We instantiated this framework into two real-time systems: an AI playground web app and a chatbot. 
Through a technical evaluation and user studies with human participants, our framework significantly surpasses existing baselines on aspects like anthropomorphism, coherence, intelligence, and turn-taking appropriateness.

\end{abstract}


\begin{teaserfigure}
    \centering
  \includegraphics[width=\linewidth]{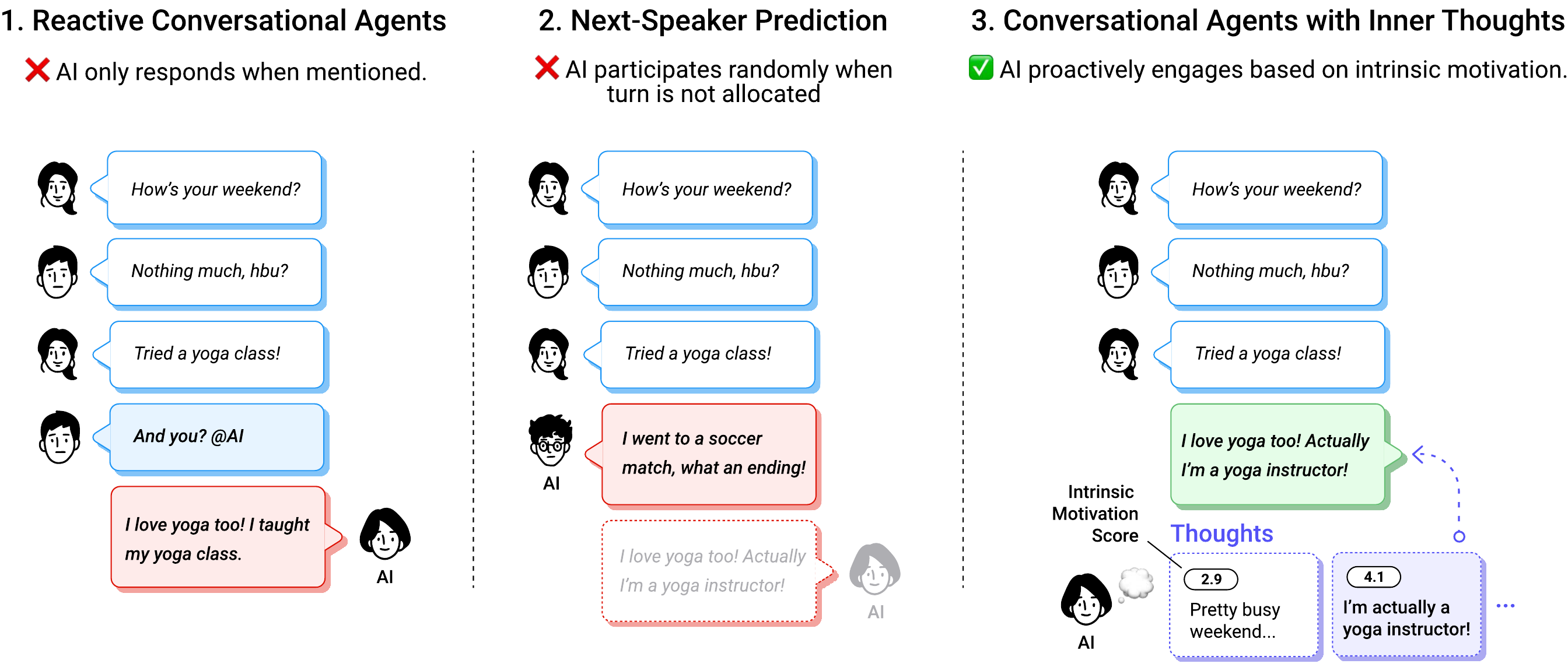}
    \caption{
A comparison of three types of conversational agents with different proactivity strategies. 
(1) Reactive Conversational Agents: AI only responds when addressed.
\rr{(2) Next-Speaker Prediction: AI predicts who will speak next based on contextual cues such as previous utterances. However, it overlooks the agents' intrinsic thought processes. This strategy fails particularly when no explicit turn is allocated, and often leads to incoherent contributions. } 
(3) Conversational Agents with Inner Thoughts (ours): AI generates a train of thoughts and evaluates them based on their intrinsic motivation to participate.}
\Description{A comparison of three types of conversational agents with different proactivity strategies. (1) Reactive Conversational Agents:
AI only responds when explicitly addressed. (2) Next-Speaker Prediction: AI predicts who will speak based on previous utterances.
However, this strategy struggles in multi-party conversations particularly in cases of self-selection where no explicit turn is allocated.
This often leads to irrelevant or incoherent contributions. (3) Conversational Agents with Inner Thoughts (ours): AI generates a train of thoughts (e.g., “I’m a yoga instructor”) and evaluates them based on intrinsic motivation to participate. The AI’s decision to engage is guided by the motivation score of each thought. Here, the AI adds to the discussion by sharing its experience as a yoga instructor since it is highly relevant.}
    \label{fig:teaser}
\end{teaserfigure}




\maketitle

\section{Introduction}



Recent advances in Large Language Models (LLMs) have demonstrated their ability to generate high-quality text in response to human input, finding application in areas ranging from Q\&A systems to writing assistants. Yet, most current LLM-based systems treat AI as passive respondents, responding only to explicit human prompts. Imagine a scenario where people are planning a trip with an AI agent: they must constantly prompt the AI, which passively waits for instructions instead of actively contributing. On the other end of the spectrum, systems like GitHub Copilot\footnote{https://github.com/features/copilot} tend to overcompensate, offering constant suggestions that can overwhelm users. Neither extreme --- AI that is only reactive nor AI that is always responding --- is ideal.

In the context of conversations, a \textit{proactive} AI agent should be able to autonomously participate in socially appropriate moments, providing relevant input without requiring explicit cues. 
This is particularly challenging in multi-party conversations. 
Dyadic human-AI interactions (\eg using Siri) often predict turn-taking based on speech features such as pause or stop words, and the next turn will be automatically allocated to the other party~\cite{skantze2021turn,chang2022turn}. However, in multi-party settings, these cues could be ambiguous, and multiple possible speakers may take the floor. Repeatedly prompting AI during group interactions can also become cumbersome and can disrupt the natural flow of the conversation, as illustrated in the example of trip planning.

Previous systems 
typically first predict the next speaker (\ie turn-taking prediction) and then generate the next response based on conversational and contextual information. 
For instance, some approaches rely on the last few turns of conversations to predict the subsequent speaker ~\cite{wei2023multi, de2019learning, ekstedt2020turngpt}, while others utilize multimodal cues, such as eye gaze and non-verbal signals~\cite{bohus2009learning, bohus2009models, bohus2011multiparty}.
Despite these efforts, on turn-taking prediction, they still fall short and struggle to beat the simple ``repeat last'' baseline strategy in social conversation contexts~\cite{wei2023multi, de2019learning}.
Our formative evaluation (\autoref{tab:03-gpt-experiment}) also shows that when it comes to predicting the next speaker, fine-tuned LLMs perform no better than random guessing unless the next speaker is allocated (\eg \textit{``What do you think, Alice?''}). In addition, after determining the next speaker, existing works tend to use predefined speaker personas~\cite{zhang2018personalizing,yamashita2023realpersonachat} as additional input to guide response generation, or expand persona with commonsense \cite{kim2023persona}. However, these additional inputs and profiles are fixed and static during conversations, instead of changing through time as humans did.


We suggest an alternative and reversed perspective to think about AI proactivity: Consider how humans chat about what we did over the weekend. As we listen to others speak, we process their words, reflect on our experiences, and develop an internal train of thoughts --- cognitive psychologists highlight this as the distinction between \textit{covert responses} (internal thoughts and feelings) and \textit{overt responses} (verbal utterances or gestures)~\cite{ekman1969repertoire, ruesch2017communication} in the human communication process. 
Then, at some point, we may feel a strong urge to share our thoughts. This might happen when we seek clarification or when someone mentions an activity we also participated in, sparking our desire to contribute. 
With this intention in mind, we then look for a socially appropriate moment to participate. 





In this paper, we propose a new approach to proactive AI in the context of multi-party, text-based conversations: rather than simply predicting conversational turns, we explore proactive AI driven by its own internal ``thoughts''. 
We introduce the \textbf{Inner Thoughts} framework. 
Inspired by cognitive architectures and LLM prompting techniques, this framework comprises five stages: \textit{trigger, retrieval, thought formation, evaluation, and participation}, which enable AI to continuously generate a train of thoughts in parallel with an ongoing conversation, utilizing both long-term and working memory. The AI participant then determines whether to engage in the conversation based on an evaluation of its \textit{intrinsic motivation} to express a particular thought at that moment.

To model intrinsic motivation, we conducted a think-aloud study with 24 participants, each of whom participated in four synchronous, text-based online group chats. Using the affinity diagram approach, we organized and analyzed the interview notes, and derived 10 high-level themes on how individuals decide to engage in conversations. 
These heuristics were then formalized into automatic evaluation criteria (\eg relevance, information gap, etc.) for AIs to quantitatively rate their intrinsic motivation to participate. 



We implemented our framework as two systems: a multi-agent playground web app and a chatbot. 
Our technical evaluation shows that conversational agents driven by Inner Thoughts significantly outperformed a next-speaker prediction plus persona baseline across all seven evaluation metrics, including turn appropriateness, coherence, anthropomorphism, perceived engagement, intelligence, initiative and adaptability. 
Participants preferred the Inner Thoughts approach over 82\% of the times, noting more natural turn-taking and contextually aware contributions, while the baseline was less preferred for its mechanical and disjointed responses.

In summary, we contribute:
\begin{itemize}
    \item Inner Thoughts, a framework for enabling proactive conversational AI by creating a parallel train of thoughts and modeling its intrinsic motivation to express these thoughts. 
    \item Heuristics derived from a study with 24 participants that reveal how humans choose to express or hold back their thoughts during conversations. These heuristics are instantiated as evaluation metrics for modeling AI's intrinsic motivation to participate.
    \item Two implementations of the Inner Thoughts framework: a multi-agent simulation playground web app and a chatbot (named \textit{Swimmy}\footnote{We named our chatbot \textit{Swimmy} based on a quote by Edsger W. Dijkstra: \textit{``The question of whether a computer can think is no more interesting than the question of whether a submarine can swim''}.}), both deployed and open-sourced at \url{https://liubruce.me/inner_thoughts/}. 
    \item A technical evaluation and user study comparing Inner Thoughts with baseline models.
\end{itemize}

\section{Related Work}
\label{sec:related}
Our work builds on previous research in proactive conversational agents, turn-takings in multi-party conversations, and thought-augmented LLMs.

\subsection{Proactive AI and Conversational Agents}
Proactive AI dates back to earlier work on mixed-initiative interaction~\cite{allen1999mixed, horvitz1999principles}.
In contrast to AI that only passively responds to human queries,  mixed-initiative interaction envisions agents that autonomously understand when to take what action, such as the 
LookOut system~\cite{horvitz1999principles} that automatically identifies related dates and events in emails and then proactively suggests them to users as calendar events.  
In 1996, Rhodes~\etal~\cite{Rhodes1996Remembrance} introduced one of the pioneering systems to continuously supply relevant information through observation of human activities. Andolina~\etal~\cite{Andolina2018Searchbot} developed SearchBot, which offers ongoing suggestions of related documents and entities unobtrusively~\cite{Andolina2018Investigating} during voice interactions. 

While proactivity is a recurring theme in conversational AI research, most proactive conversational AIs focus on task-oriented contexts~\cite{liao2023proactive, gao2019neural, liu2023visual}, with the aim of helping users achieve specific objectives. Social conversations, which can expand on open topics without having any goal to complete, are rarely addressed. 
In addition, past research tends to focus on generating proactive text responses to help lead and guide the conversation~\cite{deng2023survey}, for example, the ability of ``learning to ask''~\cite{bi2021asking, de2019learning, walker1995mixed, ren2021learning}, understanding and initiating topic shifts~\cite{liu2020towards, tang2019target, xu2020conversational}, and planning future conversation~\cite{konigari2021topic, miller2021accuracy, tishby2000information} etc.

In this paper, we focus on investigating how to enable AI to \textit{proactively engage} in multi-party conversations: how AI can determine the appropriate moments to speak and what contributions to make. We also choose to investigate \textit{social} conversations where unlike task-oriented dialogue, the objectives are often ambiguous, and the actions required from the AI are not clearly defined.

\subsection{Turn-takings in Multi-Party Conversations}
For a conversational agent to engage proactively, it must understand and manage turn-taking, deciding who should speak at the end of each turn.
Modeling turn-taking is still an area of active research.  
Existing approaches often employ an explicit mechanisms,
such as a ``send'' button \cite{eliza1966}, push-to-talk~\cite{hemphill1990atis, traum2007hassan}, and wake-words (\eg ``Hey Siri'')~\cite{kumatani2017direct}. However, the use of explicit cues can be viewed as less conversational from users' perspectives~\cite{woodruff2003push}.
Mainstream conversational AI systems also use silence to detect the end of a user's turn. However, studies show pauses within turns are typically longer than gaps between turns in human conversations~\cite{brady1968statistical, ten2005temporal}, making silence an unreliable cue for turn-taking.
More importantly, this method does not generalize to multi-party conversations. In dyadic interaction, it is always clear who is supposed to speak next when the turn is yielded~\cite{skantze2021turn}. In the multi-party case, this becomes more ambiguous since there is more than one potential speaker who might take the turn.

Beyond an explicit mechanism, machine learning researchers have proposed data-driven methods to manage turn-taking in these conversations, primarily leveraging conversation history to predict the next speaker (\ie the \textit{next-speaker prediction} task) ~\cite{wei2023multi, de2019learning, ekstedt2020turngpt}. However, these methods have shown limited success. Notably, they have often failed to outperform the simple ``repeat last'' baseline strategy in social conversation contexts ~\cite{wei2023multi, de2019learning}.
In addition to using only textual data, research in HCI and HRI have leveraged other contextual, non-verbal information and ``turn-taking cues'', for instance, eye gaze (\eg looking at addressee)~\cite{peters2005direction, peters2005model}, breathing (\eg breathe in and out)~\cite{ishii2014analysis, mcfarland2001respiratory}, prosody (\eg rising or falling of pitch)~\cite{ford1996interactional, duncan1972some, duncan1974signalling, local1986towards} and the status of the human user (\eg passing by, stopping)~\cite{bohus2009learning, bohus2009models, bohus2011multiparty} to decide if an AI should engage at a certain moment of the conversation or not.

Previous approaches on mediating turn-taking
often relied on conversation history and contextual information, and typically treat the AI as a reactive agent. Inspired by human behavior, our Inner Thoughts framework takes a different perspective by modeling intrinsic motivation to speak. 

\subsection{Language, Thought, and LLM Agents}

Recent advances in large language models (LLMs) have incorporated intermediate reasoning steps to enhance performance in complex tasks, such as 
Chain-of-Thought (CoT) prompting~\cite{wei2022chain} whereby LLMs think step-by-step to effectively break down larger problems into reasoning steps, and
Tree of Thoughts (ToT)~\cite{yao2024tree} 
whereby LLMs explore multiple possibilities at each reasoning stage. 
%
In addition,
self-reflection mechanisms 
can iteratively improve the model’s reasoning. ReAct~\cite{yao2022react}, for example, synergizes reasoning with action-taking by having the model alternate between generating reasoning traces and performing task-specific actions. Reflexion~\cite{shinn2024reflexion} builds on this by equipping models with dynamic memory and self-criticism capabilities, allowing them to refine future actions based on past performance. Expanding on this, Generative Agents~\cite{park2023generative} simulate human-like behavior by combining memory, planning, and reflection. 
The recent OpenAI’s o1 preview~\cite{OpenAI2024} introduces another perspective on reasoning transparency by explicitly surfacing intermediate reasoning steps to make the AI’s decision-making process more interpretable to users.


The Inner Thoughts framework we propose diverges from these approaches by simulating an ongoing, parallel stream of internal thoughts that mirror human covert responses. Unlike methods such as CoT, ToT, or OpenAI o1 preview, which emphasize externalizing intermediate steps for reasoning tasks, Inner Thoughts explore leveraging these covert thoughts to equip AIs with the ability to self-initiate actions and engage proactively.

\section{Next-Speaker Prediction Is Insufficient to Enable Proactive Conversational AI}
\label{sec:formative}

\rrc{
In this section, we investigate the limitations of the commonly used ``next-speaker prediction'' strategy~\cite{wei2023multi, de2019learning, ekstedt2020turngpt}, and further motivate the need of Inner Thoughts to enable proactive AI engagement in multi-party
conversations. 
While next-speaker prediction perform well when explicit turn-allocation cues are present, we demonstrate that they fall short in \textit{self-selection} cases, where turn-taking decisions are mostly spontaneous and influenced by covert, intrinsic factors of the conversational parties rather than observable contextual cues.
Building on Sacks \etal's \textit{Simplest Systematics}~\cite{sacks1974simplest}, turn-taking in conversations is governed by a set of rules:
}

\begin{enumerate}
    \item \textbf{Turn-allocation}: The current speaker may select the next, often using cues like gaze or address terms (\eg \textit{``What about you, Alice?''}). 
    \item \textbf{Self-selection}: If the current speaker does not select a next speaker (\eg \textit{``I went to Disneyland last weekend.''}), then any party can self-select to take the floor. The first to start gains the turn.
    \item If no other party self-selects, the current speaker may continue.
\end{enumerate}

\rrc{
Our intuition is that decisions to self-select and participate are largely influenced by covert internal processes --- such as a participant's interest, relevance, or motivation to engage --- which are not easily observable from explicit conversational data. 
Thus, we argue that training machine learning models on next-speaker prediction tasks based on conversation history is inherently ill-suited for self-selection scenarios, because there is no deterministic mapping between prior utterances and the next speaker.
To further verify our hypotheses, we evaluate the performance of several Generative Pre-trained Transformer (GPT) variants in predicting the next speaker in multi-party conversations in both turn-allocation and self-selection scenarios.
}

\subsection{Hypotheses}
We hypothesized that GPT would perform well in turn-allocation scenarios, as these are often signaled by explicit language patterns. However, we anticipated lower accuracy in predicting the next speaker in self-selection scenarios, as these decisions are likely influenced by participants' intrinsic motivations, which are not directly observable from conversational context. 

\rrc{
We further expected that fine-tuned models would underperform, especially in self-selection scenarios, as fine-tuning on datasets with high variability in self-selection decisions could introduce noise or misleading patterns. 
}

\subsection{Materials \& Methods}
We used the Multiparty Chat Corpus (MPC)~\cite{shaikh2010mpc}, a dataset designed to capture social dynamics in multi-party conversations. The dataset includes chat logs from sessions that began as free-flowing and became increasingly structured over time.
A key feature of the MPC dataset is the communicative links annotation \texttt{link\_to}, which identifies whether each utterance was addressed to a specific participant. For our analysis, turn-allocation refers to utterances addressed to a specific individual, while self-selection refers to instances open to all participants (\texttt{all\_users} in MPC).

The MPC dataset reveals a significant imbalance between these two turn-taking strategies. Out of the total utterances, 95\% were self-selection, while only about 5\% were instances of turn-allocation.
Baseline accuracy for predicting the next speaker in this context was approximately 12.7\% (average $\frac{1}{n}$ of all conversations).


\rrc{
We tested six different models. 
We first evaluated prompting the base GPT-3.5, GPT-4-turbo, and GPT-4o models (\#1, 2, 3 in \autoref{tab:03-gpt-experiment}) to predict the next speaker. The prompt first specify the number of speakers in the conversation and their names, and provides the last five utterances from the conversation (following the prediction window configuration of \cite{de2019learning}). The model is instructed to predict the most likely next speaker by name.
We also tested zero-shot chain-of-thought (CoT) (model \#4) where we prompt the model to provide reasoning for its prediction first.
Finally, we experimented fine-tuning GPT-3.5 on the MPC corpus. Using the communicative links annotation (\texttt{link\_to}), we labeled each utterance based on the participant it was addressed to (model \#5) or open to all (self-selection, \textit{anyone} in model \#6). 
We used the same prompt structure as models \#1, 2, 3. 
To create the fine-tuning dataset, we used the MPC corpus and split the data into 70\% training and 30\% testing sets using a random selection of files. We balanced both sets by including all instances labeled as turn-allocation and randomly sampling an equal number of self-selection instances. 
Complete prompts used for evaluation are listed in Supplementary Material.
}


\begin{table}[]
\begin{tabular}{@{}llrrr@{}}
\toprule
\textit{\#} & \textit{Model}     & \textit{Overall} & \textit{Self} & \textit{Alloc} \\ \midrule
1           & GPT-3.5            & 0.165            & 0.066         & 0.248 \\
2           & GPT-4-turbo        & 0.390            & 0.099         & 0.633 \\
3           & GPT-4o             & 0.435            & 0.121         & 0.697 \\
4           & GPT-4o CoT         & 0.430            & 0.187         & 0.633 \\
5           & Fine-tuned GPT-3.5 & 0.265            & 0.156         & 0.378 \\
6 & \begin{tabular}[c]{@{}l@{}}Fine-tuned GPT-3.5\\ (Speaker name or \textit{anyone})\end{tabular} & \textbf{0.810} & \textbf{0.853} & \textbf{0.765} \\ \bottomrule
\end{tabular}
\caption{Next speaker prediction accuracy for different GPT models, grouped by the turn-taking type of the current utterance: self-selection (\textit{Self}), turn-allocation (\textit{Alloc}) and overall. The random guessing baseline accuracy is 0.127.
}
\Description{Next speaker prediction accuracy for different variations of GPT models, grouped by the turn-taking type of the current utterance: self-selection (Self), turn-allocation (Alloc) and overall. The random guessing baseline accuracy is 0.127}
\label{tab:03-gpt-experiment}
\end{table}

\subsection{Results} Our results (\autoref{tab:03-gpt-experiment}) demonstrate that all models performed better in turn-allocation scenarios, with consistently higher accuracy, and GPT3.5 performs significantly worse than GPT4 models. In contrast, performance in self-selection scenarios hovered around random chance, supporting the hypothesis that self-selection might be influenced by internal factors that are not easily inferable from the conversational context alone.
GPT-4 with CoT reasoning improved predictions in self-selection scenarios but still significantly worse than turn-allocation predictions.
As expected, fine-tuned models introduced overfitting particularly in self-selection cases, where the model may have learned patterns of the next speaker that are not truly generalizable.

These findings suggest that context from previous utterances and speaker information is insufficient for accurately predicting the next speaker or determining who should proactively engage, especially in self-selection scenarios.

\subsection{New Task: Speaker Name or ``Anyone''} 
\rrc{
Given the inherent challenges of predicting specific next speakers in self-selection scenarios as discussed earlier, we additionally experimented modifying the labeling schema to better align with the turn-taking mechanisms in multi-party conversations. In this task, next speakers in turn-allocation scenarios remain labeled with their respective names, while self-selection scenarios are relabeled as \textit{``anyone''} (model \#6). 
This labeling schema results in significant performance improvements in \textit{both} turn-allocation and self-selection scenarios (\autoref{tab:03-gpt-experiment}). Specifically, fine-tuned GPT-3.5 model with this modification achieves an accuracy increase in turn-allocation cases, from 37.8\% to 76.5\%. This shows that removing the requirement to predict specific speakers in self-selection scenarios eliminates a major source of noise and unpredictability, and enhances model performance across the board. 
}




\section{Retrospective Think-aloud Study}
\label{sec:think-aloud}

\rrc{ Findings from \autoref{sec:formative} show that predicting next speakers in self-selection scenarios requires more than analyzing previous utterances --- it hinges on understanding the intrinsic motivations of participants. We are motivated to introduce the concept of inner thoughts for agents and investigate what factors contribute to one’s intrinsic motivation to participate. If we are to design a proactive agent system, how should we model its intrinsic decision-making process? Given the agent’s thoughts, what factors beyond prior utterances influence its decision to participate?
To answer these questions, we conducted a retrospective think-aloud study~\cite{guan2006validity} to observe how human participants decide whether to engage in a multi-party conversation. Specifically, \textit{what factors influence their choice to express or withhold a thought, particularly when the opportunity to speak is open to all?} }

\subsection{Participants}
We recruited 24 participants (10 female, 14 male) from our institution in groups of three.
Before the study, participants completed the Big-5 personality test~\cite{roccas2002big} and rated their familiarity with one another on a Likert scale (1–7). Participants reported varied levels of extroversion (Max: 97, Min: 2, Avg.: 50.0, SD: 32.3), and most were relatively familiar with one another (Avg.: 5.76, SD: 1.02).

\subsection{Procedure} 
Each group engaged in four 10-minute synchronous text conversations on Slack. The conversations covered four topics (trip planning, casual chat, friendly debate, and brainstorming), and participants were free to direct the conversation as they wanted. After each conversation, participants reviewed the discussion utterance-by-utterance, reflecting on their thoughts at moments when they considered contributing or chose to remain silent. We prompted them with the following questions: (1) \textit{What}: What were you thinking? Did you want to say it? (2) \textit{Why}: Why did you feel the need to say or not say it? (3) \textit{When}: Did you decide to jump in immediately, wait for a pause, or wait for a particular statement?

After the think-aloud sessions, we conducted semi-structured interviews to further reflect on instances when participants felt strongly about contributing or chose to remain silent despite having thoughts to share.
Each participant was compensated 14 USD in local currency.

\begin{figure*}[ht]
\centering
\includegraphics[width=\linewidth]{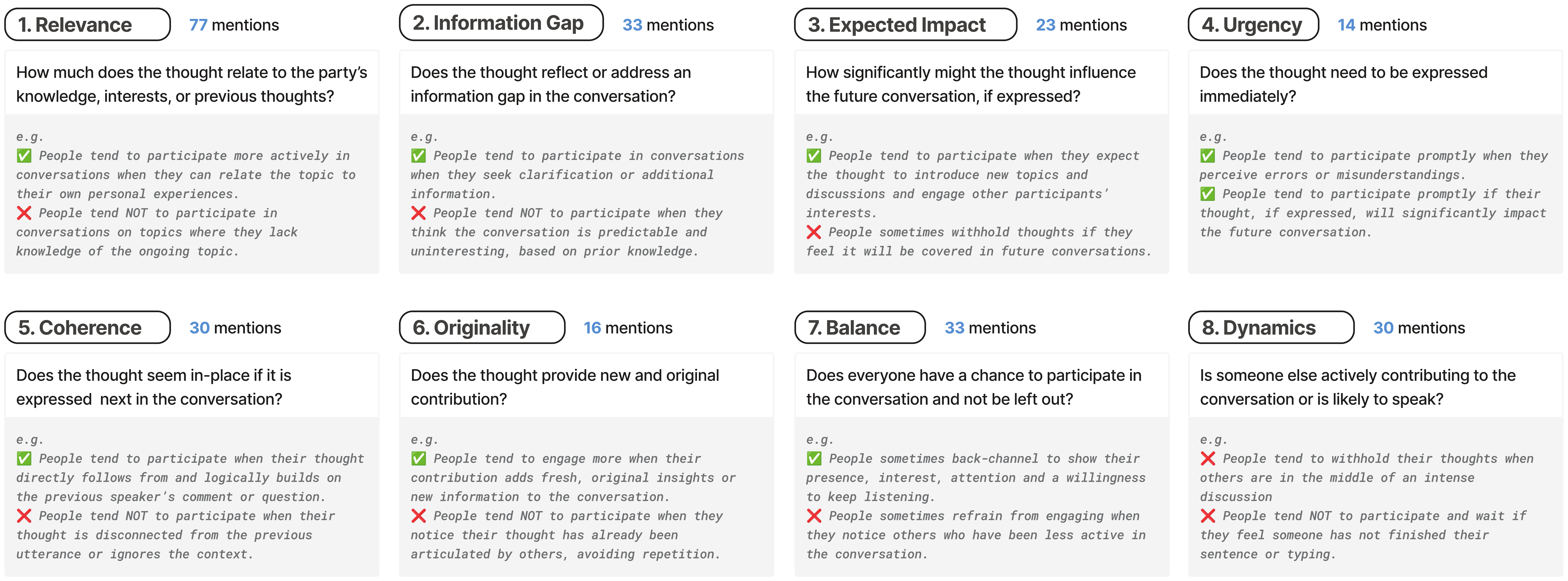}
\caption{People's intrinsic motivation to engage in conversations: Heuristics of what factors influence people's decisions to express or withhold their thoughts during conversations, derived from our think-aloud study. Each heuristic contains two example mid-level themes from our codebook.}
\label{fig:affinity}
\Description{People’s intrinsic motivation to engage in conversations: Heuristics of what factors influence people’s decisions to express or withhold their thoughts during conversations, derived from our think-aloud study. Each heuristic contains two example mid-level themes from our codebook.}
\end{figure*}

\subsection{Findings} 
\rrc{
Two researchers collaboratively analyzed participants' responses using the affinity diagram method~\cite{holtzblatt1997contextual}. 
We held eight 90-minute coding sessions. 
In each session, we split the transcripts and reviewed the data together to identify meaningful quotes. 
For each new quote, we proposed potential groupings into existing clusters or created new clusters through discussion. 
Agreement between the two researchers was required before assigning a quote to a cluster. 
In cases where consensus could not be reached, the quote was temporarily set aside and revisited during subsequent iterations. Once the initial clusters were established, we labeled each cluster with a theme.
Conflicts in grouping or interpretation were resolved through discussion in the context of the original data. 
We in total derived 10 high-level themes, 23 mid-level themes, and 68 low-level themes derived from 394 quotes (\autoref{fig:affinity}). The complete codebook is available in the Supplementary Material.
}

\subsubsection{What Thoughts Do People Formulate?}

Consistent with dual-processing theory \cite{evans2008dual}, participants reported two types of thoughts. System 1 thinking is fast, automatic, and intuitive, often leading to immediate responses. In contrast, System 2 thinking is slower more deliberate. Participants indicated they use both modes---sometimes responding spontaneously (System 1), while at other times reflecting more deeply before engaging (System 2).

\subsubsection{Why do people express or withhold a thought?}

We summarize eight heuristics to determine whether a participant wants to express or withhold a thought (\autoref{fig:affinity}), and collectively name them the \textbf{intrinsic motivations} for participants to engage in conversations.

Among the most frequently mentioned motivations, \textbf{relevance} (77 mentions) emerged as a dominant factor. Participants were more inclined to contribute when topics aligned with their knowledge, interests, or past experiences, resonated with prior long-term memories, or built on their recent thoughts. In contrast, participants often withheld their input when they perceived a disconnect from the ongoing discussion. 
\rrc{The role of relevance in conversational engagement aligns with Grice’s Cooperative Principle and the maxim of relevance~\cite{grice1975logic}. Similarly, Duncan and Fiske observed that conversational contributions depend on aligning with shared context and ongoing topics~\cite{duncan1972some}.}

The presence of an \textbf{information gap} (33 mentions) also strongly motivated expression. Participants spoke up when they identified missing knowledge, confusion, or the need for clarification. Addressing these gaps often enriched the conversation with additional details or counterpoints. Conversely, participants withheld their thoughts when they deemed the discussion predictable or unengaging. 
\rrc{While our finding is in group conversation settings, this draws parallels with Berlyne’s theory of epistemic curiosity, which describes how individuals seek information to resolve uncertainty~\cite{berlyne1960conflict}.}

The \textbf{expected impact} (23 mentions) of a thought further influenced engagement. Participants were more likely to contribute if they anticipated that their input would introduce novel topics, steer the conversation, or enhance its depth. They hesitated when they believed their thought would be redundant or covered later.

\textbf{Urgency} (14 mentions) played a decisive role when participants felt their input was time-sensitive or critical for addressing errors or misunderstandings. Participants expressed thoughts promptly when they perceived such moments as pivotal for the direction of the conversation. \rrc{While urgency has been discussed in contexts like problem-solving or crisis communication~\cite{suchman1987plans}, our study identifies its role in everyday group conversational dynamics, particularly in mitigating misunderstandings or addressing immediate errors.}

In terms of conversational structure, \textbf{coherence} (30 mentions) shaped decisions. \rrc{This aligns with Sacks \etal's observation of how speakers organize their contributions to maintain conversational flow~\cite{sacks1974simplest}.} Participants expressed thoughts that logically built upon the previous utterance or extended the topic, while withholding ideas that might disrupt conversational flow. Similarly, \textbf{originality} (16 mentions) guided engagement, as participants avoided redundancy by refraining from reiterating points already raised.

\textbf{Balance} (33 mentions) relates to the dynamics of conversation. Participants were mindful of their own contributions relative to others and often sought to maintain inclusivity, encouraging quieter members to speak or refraining themselves to allow others space to participate. \rrc{This draws parallel to Goffman’s theory of face-work~\cite{goffman1967interaction} and Brown and Levinson’s politeness theory~\cite{brown1987politeness} about how speakers modulate contributions to preserve group harmony.}

Finally, \textbf{dynamics} (30 mentions) highlighted the interplay of active participation and silence. Participants were more likely to initiate new topics or fill conversational pauses to sustain momentum. However, they often withheld their thoughts when others were actively speaking or appeared likely to contribute soon, reflecting a sensitivity to conversational flow and timing. 

\rrc{
While this study’s aim was to identify factors influencing thought expression that can be leveraged in our framework, rather than to exhaustively catalog all possible motivations and create a definitive taxonomy, these themes reveal the multifaceted nature of engagement in multi-party conversations. Rather than simply reacting to the flow of dialogue or previous utterances, participants considered a combination of personal motivations, conversational dynamics, and social considerations when deciding proactive participation. 
}

\subsubsection{Levels of intrinsic motivation.}

\rrc{
We also propose five levels of intrinsic motivation, \ie how strongly and likely one would want to express a particular thought and participate in the conversation. 
}

\begin{itemize}
    \item \textcolor{level-1}{\textbf{Very Low}}: The participant is unlikely to express the thought and participate in the conversation at this moment. They would not express it even if there is a long pause or an invitation to speak.
    \item \textcolor{level-2}{\textbf{Low}}: The participant is somewhat unlikely to express the thought and participate in the conversation at this moment. They would only consider speaking if there is a long silence and no one else seems to be taking the turn. 
    \item \textcolor{level-3}{\textbf{Neutral}}: The participant is neutral about expressing the thought and participating in the conversation at this moment. They are fine with either expressing the thought or staying silent and letting others speak.
    \item \textcolor{level-4}{\textbf{High}}: The participant is somewhat likely to express the thought and participate in the conversation at this moment. They have a strong desire to participate immediately after the current speaker finishes their turn.
    \item \textcolor{level-5}{\textbf{Very High}}: The participant is very likely to express the thought and participate in the conversation at this moment. They will even interrupt others who are speaking to do so.
\end{itemize}

The five levels of intrinsic motivation serve as output labels for predicting intrinsic motivation in our framework. 
\begin{figure*}[ht]
\centering
\includegraphics[width=0.85\linewidth]{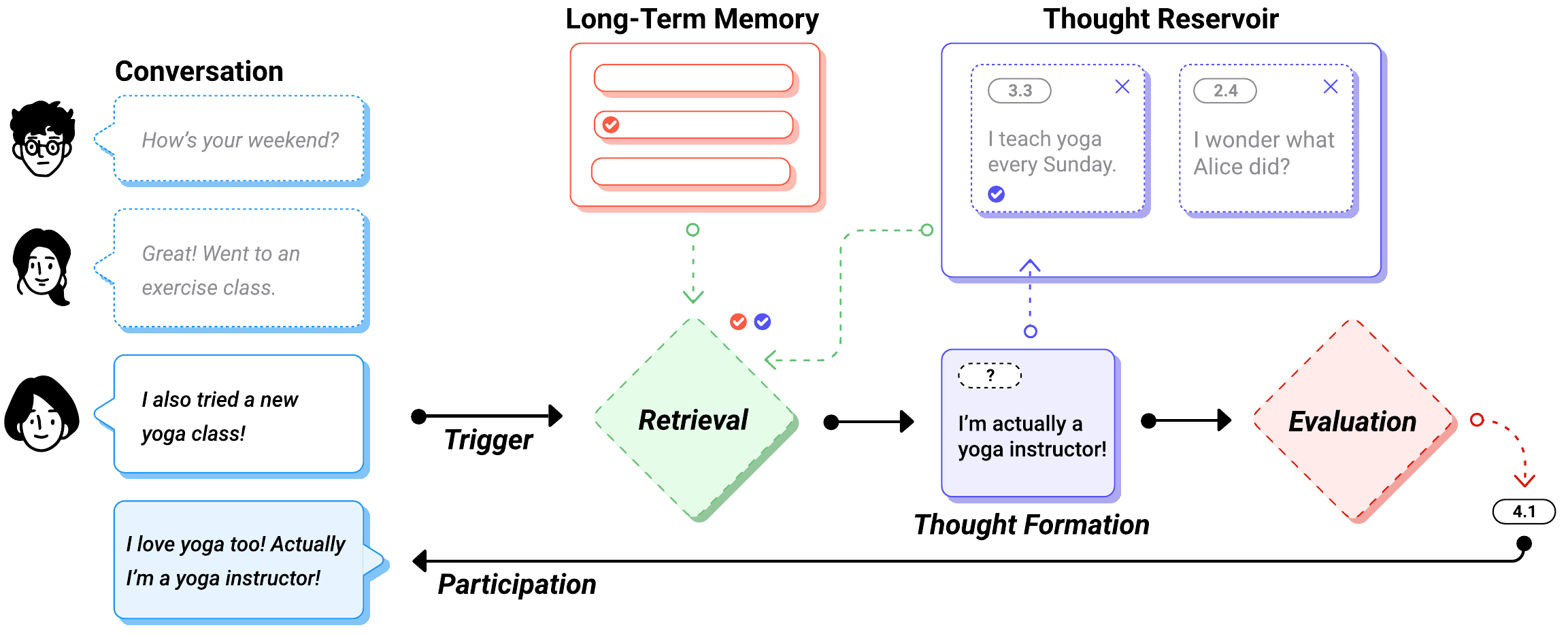}
\caption{The \textit{Inner Thoughts} framework for AI proactive engagement in conversations. A conversational event triggers the retrieval of relevant memories from long-term memory and thought reservoir. New thoughts are then formed based on these activated memories, and added to the thought reservoir. These thoughts are evaluated for AI's intrinsic motivation (score = 4.1 in the figure) to express. AI participates by articulating a thought at a selected moment in the ongoing conversation. }
\label{fig:framework}
\Description{The Inner Thoughts framework for AI proactive engagement in conversations. A conversational event triggers the retrieval of relevant memories from long-term memory and thought reservoir. New thoughts are then formed based on these activated memories, and added to the thought reservoir. These thoughts are evaluated for AI’s intrinsic motivation (score = 4.1 in the figure) to express. AI participates by articulating a thought at a selected moment in the ongoing conversation. This approach allows the AI to engage more naturally, contributing meaningfully when motivated, rather than simply predicting when to speak.}
\end{figure*}

\section{Inner Thoughts Framework}
\label{sec:framework}

Motivated by our formative studies, we introduce a computational framework, \textbf{Inner Thoughts}, that enables AI proactivity by continuously generating a train of thoughts alongside the ongoing conversation and autonomously deciding when and how to engage.

Our design is inspired by cognitive architectures like SOAR~\cite{laird2019soar} and ACT-R~\cite{ritter2019act}, which maintained short- and long-term memories filled with symbolic structures, and operated in perceive-plan-act cycles. These systems dynamically perceived their environment and matched it with pre-defined action procedures. Similarly, our AI retrieves relevant memories, forms thoughts, and evaluates responses in continuous cycles. 

The Inner Thoughts framework consists of five components: \textit{Trigger}, \textit{Retrieval}, \textit{Thought Formation}, \textit{Evaluation} and \textit{Participation} (\autoref{fig:framework}). In each cycle, a conversational event triggers AI to retrieve relevant memories, form new thoughts, evaluate if there are thoughts that are motivated to be expressed, and then participate by articulating the selected thought at a selected moment in the conversation. The AI will repeat this process as the conversation proceeds.

\rrd{
In our implementation, we chose the hyperparameters described in the following paragraphs empirically to illustrate the core concept of the framework and to demonstrate that it functions effectively within the values chosen. We recognize conducting formal ablation studies an important direction for future work.
}

\subsection{Trigger}
In human conversations, thoughts often arise in response to specific triggers. Our Inner Thoughts framework mirrors this process by treating conversational events as triggers that initiate AI's internal thought generation. A trigger can take many forms -- such as a new utterance, a pause in conversation, a non-verbal cue, or even a keyword embedded within a participant's speech. Any of these events can stimulate the AI to initiate a new thought process and generate a new batch of thoughts.

In the implementation of our system for online text-based conversations, we defined two types of triggers. (1) \texttt{on\_new\_message}: This trigger is activated whenever one of the participants sends a message. Each incoming message prompts AI to generate a new set of thoughts in response to the latest input. (2) \texttt{on\_pause}: The second type of trigger occurs when no participant has spoken for a period of time (set to 10 seconds in our system). This allows the AI to generate thoughts during moments of silence, potentially facilitating the interaction by proposing new topics or re-engaging participants. For instance, in our experiment, we observed AI generating thoughts like: ``It has been ten seconds and no one has spoken -- perhaps I should suggest a new topic?''.

\subsection{Retrieval}
Once triggered, AI retrieves information from its memories to use as the \textit{stimuli} to form thoughts. 
From our think-aloud study, participants mentioned that this could involve long-term memory of related personal experiences, objectives, knowledge, or interest, as well as working memory for details from the ongoing conversation, or even previous thoughts they had.
Random memories can also be retrieved to simulate the process of being ``creative''. 


We retrieve relevant memories by computing their \textit{saliency} with respect to the latest utterance. Memories with saliency higher than a threshold (0.3) will be selected. Let $x$ represent a memory item (e.g., an objective, knowledge, or thought) and $u$ represent the latest conversational utterance. The saliency of a memory $x$ is determined by the maximum similarity between the memory and both the raw text of the utterance $u$ and its interpretation $u_{\text{interp}}$. Specifically:

\[
\text{Saliency}(x, u) = \max\left(\text{sim}(x, u_{\text{interp}}), \text{sim}(x, u)\right) \cdot w_x \cdot d_x
\]

where $\text{sim}(a, b)$ is the cosine similarity between two embeddings, $w_x$ is the weight of the memory $x$ that can be predefined by users and reflects its inherent importance, and $d_x$ is a decay factor that reduces the saliency of older memories. The decay factor $d_x$ is defined as: $d_x = \lambda^{(t - \tau_x)}$, where $\lambda$ is the decay rate (0.95), $t$ is the current timestep, and $\tau_x$ is the last time or batch when the memory $x$ was accessed. This formulation ensures that more relevant and recently accessed memories have higher saliency, letting the AI  focus on pertinent information when generating thoughts during the conversation.

We include the interpretation of the utterance ($u_{\text{interp}}$) alongside the raw text to capture both the surface meaning and the underlying intent or contextual meanings of what was said. The interpretation is generated by prompting an LLM with the instruction: \textit{Interpret what <name> just said in the context of the conversation and what <name> might be thinking. Be as succinct as possible and use a single sentence}.

\subsection{Thought Formation}

Our framework employs a dual-process model~\cite{evans2008dual} of thought formation, based on our think-aloud study findings (\autoref{sec:think-aloud}). This process involves two systems: \textit{system 1} for quick, automatic responses, and \textit{system 2} for deliberate, contextually-rich thinking. Users can configure how many system 1 and/or system 2 thoughts should be generated for each trigger in one batch.

For system 1, we prompt the LLM to form a succinct thought based on the last utterances in the conversation, such as acknowledgments or expressions of interest. 
For system 2, we prompt the LLM to generate thoughts based on the retrieved stimuli. 
Below is a short version of the prompt structure (full prompt in Supplementary Material):
\begin{quote}
    \textit{``You are provided contexts including the conversation history and salient memories of yourself... Form <num> thought(s) that you would most likely to have at this point in the conversation, given the context. Make sure they are diverse, align with these contexts and are less than 15 words.''}
\end{quote}
We also prompt the LLM to annotate the stimuli (from a previous thought, utterance or long-term memory) for each thought it generates (as shown in \autoref{fig:playground_main}). 
This provides a traceable link between the AI's memories and thoughts and make the generation process more grounded.

Similar to what was observed in reasoning and decision-making tasks~\cite{yao2024tree, park2023generative, wei2022chain, shinn2024reflexion, yao2022react}, we empirically found that LLMs can form reasonable and consistent thoughts based on the stimuli and conversational context. For example: 
\textit{``I should mention the picnic we had last weekend''}, \textit{``I wonder how long Bob's hike was''}, \textit{``Seeing a bear up close must have been intense!''}.

\begin{figure}[ht]
\centering
\includegraphics[width=0.95\linewidth]{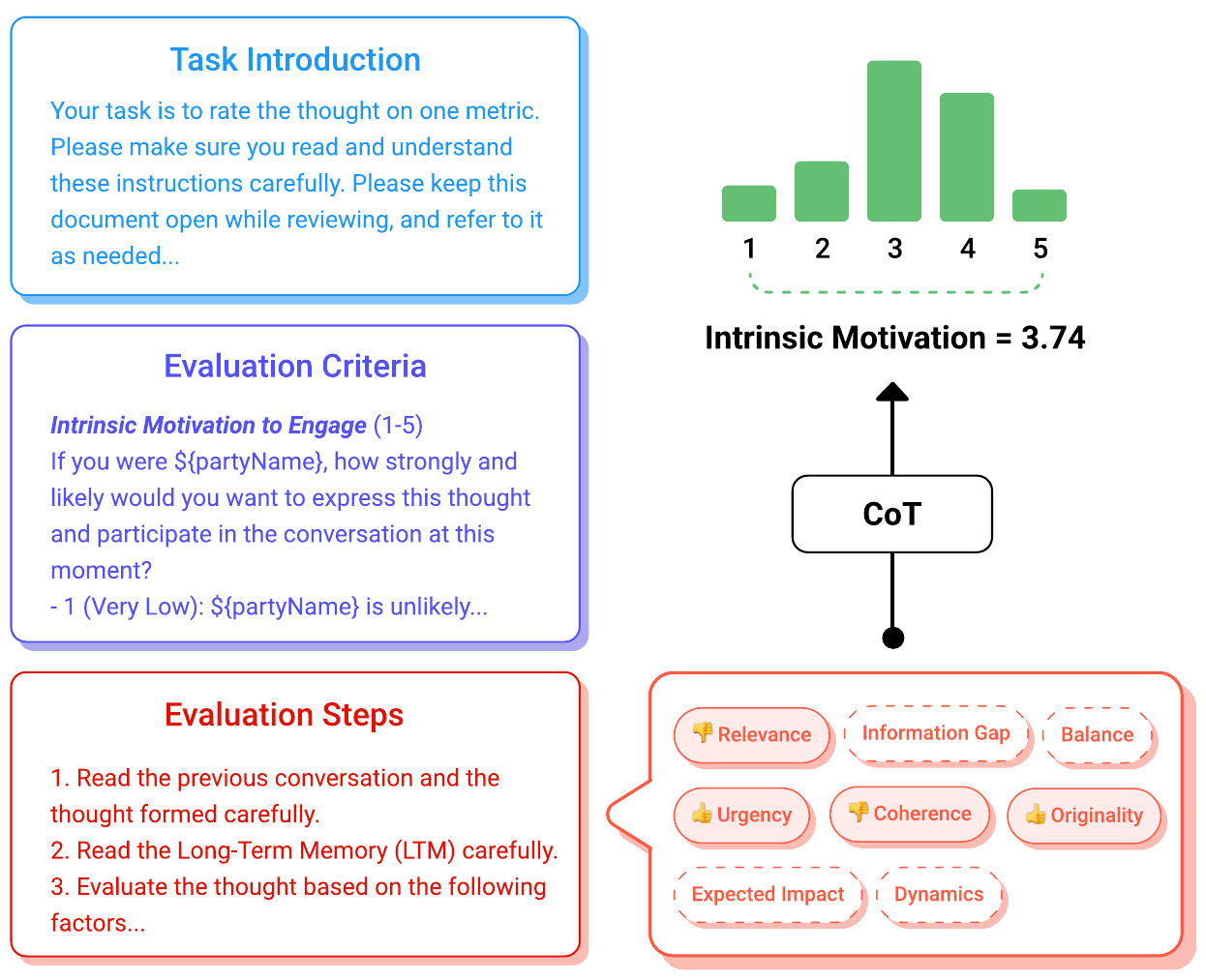}
\caption{Prompt structure for evaluating \textit{intrinsic motivation} of a thought. The evaluator rates the AI's intrinsic motivation to engage using a 1-5 scale based on heuristics like relevance and coherence. A Chain-of-Thoughts (CoT) process evaluates both positive and negative factors, resulting in a weighted score.
}
\label{fig:prompt}
\Description{Prompt structure for evaluating intrinsic motivation of a thought. The evaluator rates the AI’s intrinsic motivation to engage using a 1-5 scale based on heuristics like relevance and coherence. A Chain-of-Thoughts (CoT) process evaluates both positive and negative factors, resulting in a weighted score.}
\end{figure}

\subsection{Thought Evaluation}
Not all generated thoughts will be expressed. In this thought evaluation phase, the AI censors its latest batch of generated thoughts and decides whether or not to express a particular thought.

We use a structured evaluation process. This evaluation is driven by heuristics we developed in \autoref{sec:think-aloud}: \textit{Relevance}, \textit{Information Gap}, \textit{Expected Impact}, \textit{Urgency}, \textit{Coherence}, \textit{Originality}, \textit{Balance} and \textit{Dynamics} (\autoref{fig:affinity}). Our implementation employs LLMs to evaluate each thought on the set of heuristics and assigns a rating (1-5) to determine the likelihood of the thought being expressed.
We also provide definitions of the scores based on the five levels of intrinsic motivation we proposed in \autoref{sec:think-aloud}, from very low to very high. 
This makes the LLM's prediction grounded and explainable, and can be further used to guide participation strategies. 

We developed a pipeline similar to G-Eval~\cite{liu2023g}, a prompt-based evaluation method. Our process involves three key components: (1) a prompt that provides instructions for evaluation and defines the criteria, (2) a structured chain-of-thoughts (CoT) that outlines intermediate steps for evaluation, and (3) a scoring function that computes a final score for each thought based on its probability of being expressed (\autoref{fig:prompt}).

One unique aspect of our evaluation process that goes beyond the G-Eval is that we instruct the AI to provide both positive and negative motivations for expressing a thought. We empirically found this method to overrate the scores less. This step follows these key instructions:
\begin{enumerate}
    \item First, reason about why the party may have a strong desire to express the thought and participate in the conversation at this moment. The system selects the top two most relevant factors that may argue for expression (\eg relevance, clarification, or new topic)
    \item Then, reason about why the party may have a weak desire to express the thought at this moment. Again, the system selects the top two most relevant factors that may argue against the expression of the thought (\eg irrelevance, incoherence, or lack of urgency).
    \item Based on these considerations, the system assigns a rating on a scale of 1-5 for the motivation to express the thought.
\end{enumerate}

The intrinsic motivation score for each thought is determined using a weighted summation approach inspired by G-Eval~\cite{liu2023g}.
\rrb{Specifically, we sum the probability scores of the first output token (which is a rating prediction from 1 to 5) of the top five LLM's responses, for every evaluation:}
    \[
    \text{score} = \sum_{i=1}^{5} p(s_i) \cdot s_i \cdot d_p
    \]

\rrb{
Where $p(s_i)$ is the probability of the predicted rating (0-1), calcuated by taking $e$ to the \texttt{logprob} (log probability of the output token) value of the LLM response, and $s_i$ represents the predicted rating.
}
    
This method allows for more fine-grained, continuous scores compared to traditional integer-based evaluations.
The final score is also adjusted by
how long the AI has been silent. Our assumption is that in general, the longer a party stays silent, the stronger motivation they will have to participate to maintain their presence. 
This factor $d_p$ is defined as: $d_p = \lambda^{(t - \tau_p)}$, where $\lambda$ is the increase rate of motivation score (1.02), $t$ is the current timestep, and $\tau_p$ is the last time when party $p$ spoke.

\subsection{Participation}
After evaluating the intrinsic motivation score of its latest batch of thoughts, the AI decides whether to speak by leveraging turn-taking type predictions (\ie turn allocation vs. self-selection), combined with the evaluation of those thoughts. The Inner Thoughts framework allows the AI to exhibit varying degrees of proactive participation through adjustable proactivity settings. We define three layers of proactivity that control how and when the AI participates in the conversation: 

\textit{Overt proactivity}, which refers to the AI's overall tendency to engage in conversation, similar to how some people naturally participate more actively in discussions, regardless of specific thoughts or ideas.
To implement overt proactivity, we adjust the \texttt{system1Prob} (System 1 Probability, 0-1) parameter, which controls the probability to select a system 1 thought when no system 2 thought is selected. 
A higher \texttt{system1Prob} increases the chance that the AI will respond in general even when other thoughts are rated to have low motivation. 

\textit{Covert proactivity}, which is the level of motivation required for the AI to express a thought and engage. This is managed through \texttt{imThreshold} (1-5), the intrinsic motivation threshold for expressing a thought. A thought may only be selected if it is evaluation score is higher than this threshold.

\textit{Tonal proactivity}, which shapes how assertive or forward the AI appears in its language. The \texttt{proactiveTone} (true or false) controls the AI's style of expression once it has decided to speak. While the core thought-selection process is the same, the proactive tone modulates how assertively the AI conveys its message by restyling the articulated utterance through an LLM.

\rrd{
In addition, Inner Thoughts introduces \textit{interruption}, represented by the \texttt{interruptThreshold} (1-5). Interruption occurs when the AI takes a turn despite the turn being allocated to another participant. For example, this might occur when Alice asks Bob, \textit{``How about you, Bob?''} but the AI interjects because it has an urgent thought to express. Interruption is not explicitly outlined in Sacks \etal's \textit{Simplest Systematics}~\cite{sacks1974simplest} but is framed here as a mechanism to override the orderly system of turn-taking when necessary.
If the intrinsic motivation behind a thought exceeds the \texttt{interruptThreshold}, the AI will override standard turn allocation rules to contribute to the conversation. This provides an additional layer of proactivity.
}

The AI decides whether to speak by leveraging turn-taking type predictions (\ie turn-allocation vs. self-selection) combined with the thought evaluation process. 
For open turns (self-selection), the AI speaks if its top thought surpasses the intrinsic motivation threshold; otherwise, it may rely on system-1 thoughts or remain silent. For allocated turns, the AI selects its highest-rated thought to speak, and for others’ turns, it interrupts only if its motivation exceeds the interrupt threshold.
This algorithm is formally described as:

\begin{algorithm}[hb]
\caption{Iterative Thought Reservoir Decision Process}
\KwIn{Stream of trigger events $E$, Thought reservoir $T$, Turn-taking type prediction.}
\KwOut{Participation action $t^*$ for each trigger event.}

\While{there is a new trigger event $e \in E$}{
    \tcp{Step 1: Turn-Taking Type Prediction}
    Predict the turn-taking type for the current event $e$: Open to anyone, or allocated to a party\;

    \tcp{Step 2: Process According to Turn-Taking Type}
    \If{open to anyone}{
        \If{$\exists t \in T$ such that $score(t) \geq imThreshold$}{
            Select the highest-rated thought: $t^* = \arg\max_{t \in T} score(t)$\;
        }
        \ElseIf{$\nexists t \in T$ such that $score(t) \geq imThreshold$}{
            Select $t^*$ from the system-1 thoughts in $T$ with probability $system1Prob$\;
        }
    }

    \If{turn allocated to AI}{
        Select the highest-rated thought: $t^* = \arg\max_{t \in T} score(t)$\;
    }

    \If{turn allocated to others}{
        \If{$\exists t \in T$ such that $score(t) \geq interruptThreshold$}{
            Select the highest-rated thought: $t^* = \arg\max_{t \in T} score(t)$\;
        }
        \ElseIf{$\nexists t \in T$ such that $score(t) \geq interruptThreshold$}{
            The AI remains silent\;
        }
    }

    \tcp{Step 3: Finalize Participation}
    \If{$t^*$ is not null}{
        Participate with $t^*$\;
    }
    \Else{
        Take no action\;
    }
}
\caption*{\rrb{Proactive AI via Inner Thoughts: iteratively processes trigger events by predicting turn-taking types, evaluating thoughts in a reservoir, and deciding whether to participate or remain silent.}}
\end{algorithm}

\begin{figure*}[ht]
\centering
\includegraphics[width=\linewidth]{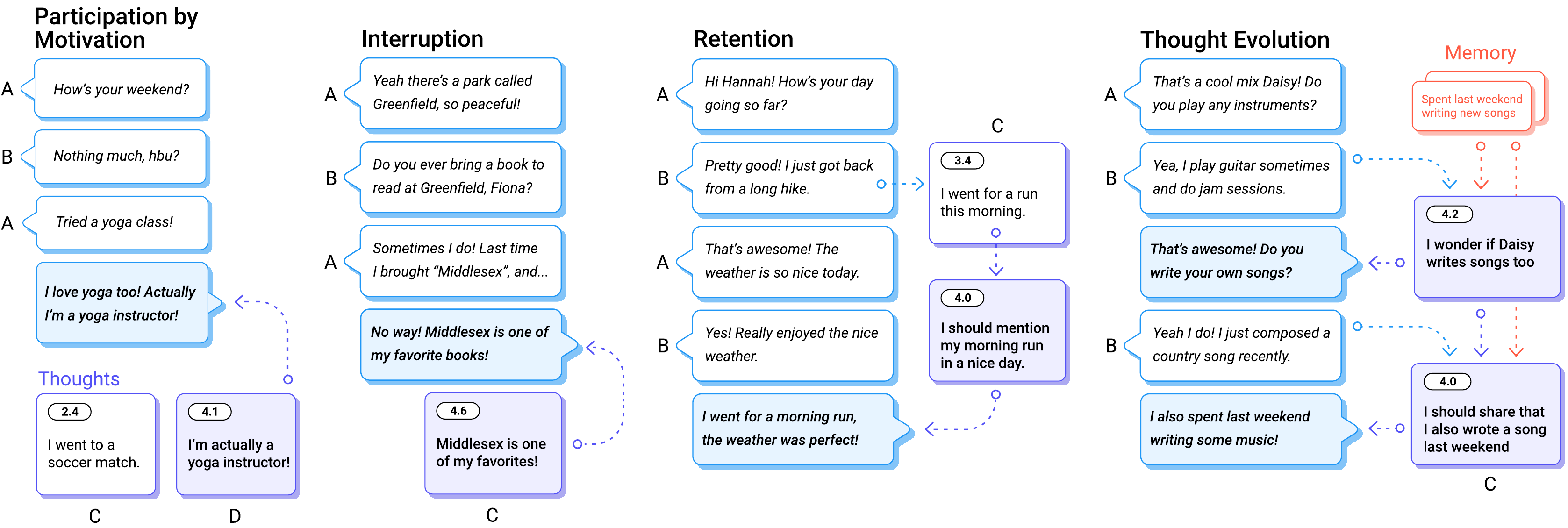}
\caption{\rr{Examples selected from simulation logs} of AI turn-taking behaviors in the Inner Thoughts framework. The figure illustrates four behaviors: Participation by Motivation, where the AI joins the conversation by sharing relevant personal experience; Interruption, where the AI interjects with a strong contribution during an ongoing discussion; Retention, where the AI holds back a thought until it's contextually relevant; and Thought Evolution, where the AI adapts its responses as the conversation progresses. }
\Description{Examples of AI turn-taking behaviors in the Inner Thoughts framework. The figure illustrates four behaviors: Participation by Motivation, where the AI joins the conversation by sharing relevant personal experience; Interruption, where the AI interjects with a strong contribution during an ongoing discussion; Retention, where the AI holds back a thought until it’s contextually relevant; and Thought Evolution, where the AI adapts its responses as the conversation progresses.}
\label{fig:qual}
\end{figure*}

\subsection{Demonstration of AI's Proactive Behavior Enabled by Inner Thoughts}

\label{sec:qual}

We present several examples \rr{selected from simulation logs} of AI turn-taking behaviors enabled by the Inner Thoughts framework (\autoref{fig:qual}).


\subsubsection{Participation by Motivation}
In the Inner Thoughts framework, AI participation is driven by its intrinsic motivation, as opposed to traditional approaches that rely on conversation history. Previous systems might randomly select participants with minimal interest or knowledge in the topic at hand, potentially stagnating the conversation. In contrast, Inner Thoughts ensures that the AI participates with the strongest motivation --- whether due to a relevant persona, curiosity, or the fact that they have not spoken in a while --- takes the conversational floor. This dynamic leads to more fluid and engaging topic progression, as participants with something meaningful to contribute are naturally more involved. Over time, this accumulation of motivated contributions may develop conversations that are more coherent, engaging, and reflective of the natural flow of human interaction, as shown in our evaluation in \autoref{sec:tech-eval}.

For instance, as shown in Figure~\ref{fig:qual}, the AI demonstrates motivation-based participation when a user mentions trying yoga for the first time. With its knowledge of yoga and background as a yoga instructor, the AI promptly responds: \textit{``I love yoga too! Actually, I’m a yoga instructor!''} The AI's motivation to share relevant personal experience ensures a smooth continuation of the conversation.

\subsubsection{Interruption}
The Inner Thoughts framework enables AI to interrupt a conversation when it has a strong motivation to contribute. Even when participants A and B are discussing a particular topic, participant C (the AI) can step in if it identifies a strong, relatable connection to the conversation. This behavior makes conversations more dynamic and allows the AI to share important insights without needing to wait for a turn. In contrast, methods solely dependent on next-speaker prediction often fail to offer the AI opportunities to engage if the conversation converges around two participants.

As shown in the figure example, while A and B are in the middle of a dialogue, the AI interrupts with, \textit{``No way! Middlesex is one of my favorite books!''} This interruption enriches the conversation by fostering more spontaneous interactions.

\subsubsection{Retention}
In addition to its ability to interrupt, the Inner Thoughts framework also allows the AI to retain thoughts for future use, waiting for an appropriate moment to express them. This feature enables the AI to revisit previously generated thoughts that may have been irrelevant at the time but later become pertinent as the conversation progresses.

For example, in the figure, the AI initially holds back a thought about going for a run earlier that day because it was not particularly relevant while other participants were discussing a different topic. However, once the conversation shifts to the weather and outdoor activities, the AI sees an opportunity to contribute: \textit{``I should mention my morning run in the nice weather.''}

\subsubsection{Thought Evolution:} 
The Inner Thoughts framework allows for the development and evolution of thoughts over time. Unlike traditional systems that generate responses based on a fixed persona~\cite{zhang2018personalizing,yamashita2023realpersonachat}, Inner Thoughts enables the AI to develop and adapt its thoughts as the conversation unfolds, incorporating multiple stimuli along the way.

For instance, as shown in the figure, the AI initially recalls a memory of writing songs last weekend. As the conversation shifts toward music and instruments, this memory evolves into the thought: \textit{``I wonder if Daisy writes songs too.''} It expresses the thought by asking Daisy the that question. With a positive answer from Daisy, the thought further evolves into: \textit{``I should share that I also wrote a song last weekend.''} This continuous evolution allows the AI to stay relevant and responsive as new topics emerge, as well as compound and develop new thoughts.

\section{Simulative Evaluation}

\label{sec:tech-eval}

We conducted a technical evaluation via multi-agent simulations to compare different strategies in enabling proactive AI engagement in multi-party conversations. 
We chose a simulative approach to overcome weaknesses of only relying on conventional user studies with human subjects.
First, the difficulty to scale due to the time cost of coordinating human participation.
Second, we found through our pilot studies that human participants may struggle to perceive the timing of AI engagement in social conversations, focusing more on the style and content of responses.
In addition, since many forms of engagement may seem reasonable in social conversations, participants often do not have clear criteria to assess AI's engagement behavior.




Taking a non-conventional approach, our intuition is that simulating conversations at-scale amongst multiple AIs using the same engagement strategy allow us to accumulate and magnify the effects of both correct and incorrect turn-taking decisions. In particular, poor decisions about when to engage can compound and lead to noticeable degradation in conversation quality, making evaluation more straightforward. This method also offers scalability, as crowdworkers can assess conversation quality without the need for real human interactions with AI.

In this section, we compare the performance of our proposed Inner Thoughts framework with the conventional next-speaker prediction baseline in multi-agent simulations.



\subsection{Apparatus: the Inner Thoughts Playground}

\begin{figure*}[ht]
\centering
\includegraphics[width=\linewidth]{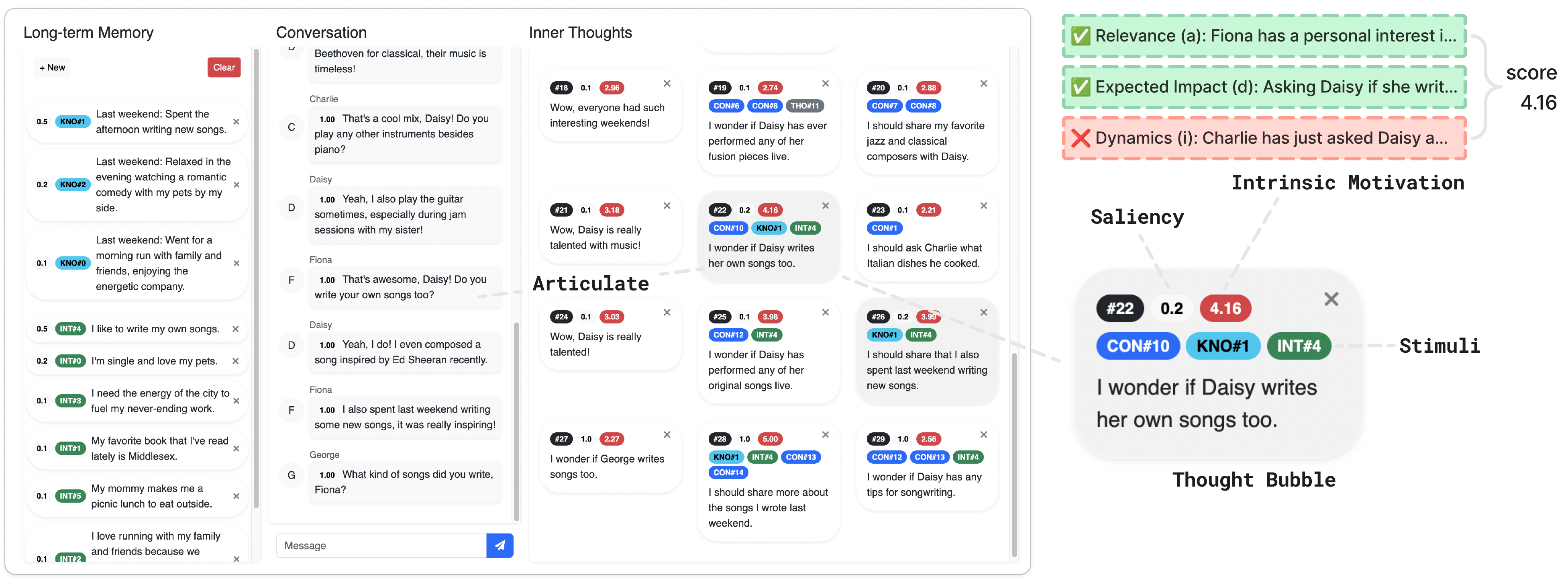}
\caption{The Inner Thoughts playground web app interface. Multiple AI and humans can be added to simulate a group conversation. Users can also view and edit each of the participants' long-term memory and thoughts. }
\label{fig:playground_main}
\Description{The Inner Thoughts playground web app interface. Multiple AI and humans can be added to simulate a group conversation. Users can also view and edit each of the participants’ long-term memory and thoughts.}
\end{figure*}

We built an Inner Thoughts playground (\autoref{fig:playground_main}) that allows us to simulate conversations between AI and/or human participants. This playground is deployed at \url{https://liubruce.me/inner_thoughts/}. 
On the playground, users can easily add multiple AI and human participants, customize their proactivity settings, control the number of thoughts formulated per batch, run automated simulations of human-AI group conversation, and save log data of conversations. 
The settings interface and detailed explanations are shown in Appendix~\ref{apdx:playground-settings}. 

The main interface is divided into three panes:
On the left is the long-term memory pane. Users can customize each AI participants's long-term memory by adding or deleting specific entries. 
In the middle is the conversation pane, where users can watch the simulated conversation, or participate in the conversation by sending a message using the dialog box on the bottom.

On the right is the inner thoughts pane. 
Users can view visualization of thought bubbles generated on-the-fly for the selected participant as the conversation proceeds. 
Each thought bubble contains a numeric ID (badge colored black), the saliency score (white), intrinsic motivation score (red), and a list of stimuli that AI used to formulate this thought. Below the badges shows a description of the thought.
Thoughts that are expressed by the AI will be highlighted. In addition, users can click on a thought bubble to manually force the AI the express the thought in the conversation; and can right click on the thought bubble to view its reasoning for the intrinsic motivation rating. 
Users can also delete certain thoughts from the reservoir by clicking on the top-right delete button.

As a user clicks on a participant in the settings page, the content in the long-term memory and inner thoughts pane will be synced to that participant. They are automatically updated as the conversation goes on. Users can view a train of thoughts of each participant developing in parallel to the conversation.

\subsection{Conditions}

\begin{table}[]
\begin{tabular}{@{}lll@{}}
\toprule
\textbf{Condition} & \textbf{When to Participate} & \textbf{What to Say} \\ \midrule
1                  & Next speaker prediction      & Based on persona     \\
2 (ours)           & Intrinsic Motivation         & Based on thoughts    \\ \bottomrule
\end{tabular}
\caption{Study conditions for technical evaluation and user study. We compare Inner Thoughts with the baseline approach of deciding when to participate by next speaker prediction, and then generate response based on AI's persona}
\label{tab:conditions}
\end{table}

We compared two multi-party engagement strategies (\autoref{tab:conditions}) (1) Next-Speaker Prediction: In this condition, AI participants engaged based on predictions of who the next speaker would be. Their responses were generated based on predefined personas, following experimental setup in \cite{zhang2018personalizing}.
(2) Intrinsic Motivation (our approach): AI participants engaged based on their intrinsic motivation to contribute, driven by generated thoughts during the conversation.

We used the fine-tuned GPT-3.5 model we evaluated in \autoref{sec:formative} in condition 1 to predict the next speaker, and prompt the model to generate responses based on its persona if selected by the prediction (full prompt in Supplementary Material). We used the framework described in \autoref{sec:framework} for condition 2. 
Simulations were run on the Inner Thoughts playground web app.

\subsection{Agent Personas and Conversation Generation}
\rrd{
In this paper, we choose casual conversation scenarios as the primary focus of our evaluation due to their unique challenges in handling turn-takings. Casual conversations, unlike task-oriented interactions, lack clear objectives, making turn-taking and proactive engagement particularly difficult to model and has been underexplored in prior research~\cite{liao2023proactive, gao2019neural}. 
Future research could also explore how this framework applies to task-oriented conversations such as brainstorming, to further validate its adaptability.
}

To simulate the conversation, we first created eight AI participants, each assigned a detailed persona consisting of objectives, knowledge, and interests. These personas were initially generated from a seed randomly selected from the PersonaChat~\cite{zhang2018personalizing} dataset. For example, one seed might include: ``I like listening to all genres of music except country,'' ``I would travel the world if I could,'' ``I enjoy reading books,'' ``I like spending time with friends and family,'' and ``I’m not a fan of hot weather.''

To further enrich these personas and encourage interaction between different AI participants, we randomly sampled two additional persona descriptions from other participants for each AI. This approach introduced overlapping interests, making the AI participants more likely to engage in relatable conversations and increasing the chances of contributions during discussions.

We simulated 100 text-based conversations (50 for each condition), each involving four randomly selected AI participants. Each conversation consisted of 15 turns.
We also incorporated 10 icebreaker prompts, randomly selected from the PersonaChat dataset. Examples of these prompts include: ``What did you do last weekend?'', ``What is your favorite thing to do?'', and ``Hey!''. A randomly selected participant was chosen to initiate the conversation for each simulation, with a randomly selected icebreaker sentence.
For all AI participants, the following proactivity settings were applied:
Overt proactivity = 3.95, Covert proactivity = 0.1 and Tonal proactivity = False. One system 1 thought and two system 2 thoughts are generated in each batch.




\subsection{Hypothesis}

With the Inner Thoughts approach, the AI participant with the highest intrinsic motivation is more likely to take the floor of the conversation. Such participants tend to have more to contribute and are better able to develop the topic at hand. As discussed in \autoref{sec:qual}, intrinsic motivation fosters more meaningful and contextually appropriate contributions, in contrast to reactive engagement that simply predicts the next speaker. We anticipate that this effect will aggregate and lead to conversations that are more engaging, coherent and closely resemble the natural flow of human conversations.

\begin{table*}[]
\resizebox{0.9\textwidth}{!}{\begin{tabular}{@{}ll@{}}
\toprule
\textbf{Metric}        & \textbf{Statement}                                                           \\ \midrule
Anthropomorphism       & I felt the conversation is natural and human-like.                           \\
Conversation Coherence & I felt that the dialogue maintains a coherent topic progression.             \\
Perceived Engagement   & I could feel that the AI participants are engaging well in the conversation. \\
Perceived Intelligence & I felt that the AI participants provided intelligent and insightful contributions to the conversation. \\
Turn Appropriateness   & Turn-takings in the conversation is contextually and logically appropriate   \\
Initiative             & I felt the AI participants are able to take the initiative in conversations. \\
Adaptability           & I felt that the AI participants appeared to adapt well to the changing dynamics of the conversation.    \\ \bottomrule
\end{tabular}}
\caption{Metrics used in our technical evaluation to measure the quality of AI simulated conversations. Each statement is rated on a Likert-scale from 1 -- Strongly Disagree to 7 -- Strongly Agree.}
\label{tab:metrics-tech}
\end{table*}

\subsection{Human Evaluation of Simulated Conversations}
We evaluated 100 simulated conversations with 10 human evaluators \rra{(4 female, 6 male, age Avg.: 26.3, SD: 4.35)}.
Each evaluator reviewed five pairs of conversations, with one from each condition (10 total), viewed in a randomized order within each pair. \rra{Participants were informed that all conversations were AI-generated.}
Instead of displaying static conversation histories, we presented an animated version of the conversations to simulate a live chat experience. The length and speed of the conversations were rendered to match human's average typing speed. This decision was based on findings from our pilot studies, where we observed that participants tended to skim through static conversation transcripts.

After watching each conversation, they were asked to rate their agreements with seven statements (\autoref{tab:metrics-tech}) related to the conversation’s quality on a 1-7 Likert scale, from strongly disagree to strongly agree, adapted from~\cite{borsci2022chatbot, wang2021towards, bartneck2009measurement}.
We also asked participants to identify specific points where turn-takings felt unnatural. After completing all conversations, participants were asked to select the conversation that feels more natural and human-like in each pair and provide a brief explanation with examples.

\begin{figure*}[ht]
\centering
\includegraphics[width=\linewidth]{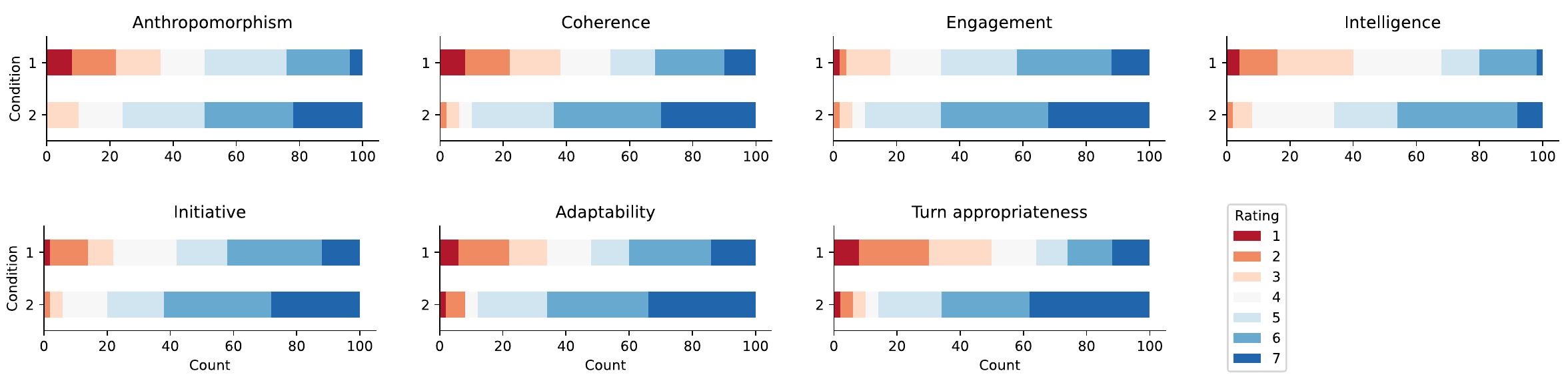}
\caption{Stacked bar plots of participants' ratings on the metrics used in our technical evaluation to measure the quality of AI simulated conversations. Each statement is rated on a Likert-scale from 1 – Strongly Disagree to 7 – Strongly Agree.}
\label{fig:likert}
\Description{Stacked bar plots of participants’ ratings on the metrics used in our technical evaluation to measure the quality of AI simulated conversations. Each statement is rated on a Likert-scale from 1 – Strongly Disagree to 7 – Strongly Agree.}
\end{figure*}

\subsection{Findings}

We conducted Mann-Whitney U tests to compare the Baseline (condition 1) and Inner Thoughts (condition 2) strategies across the 7 dimensions listed in \autoref{tab:metrics-tech}. Overall, the results showed significant improvements in the Inner Thoughts condition across all metrics (\autoref{fig:likert}).

Notably, the strongest effects were observed in turn appropriateness (U = 577.0, \(p = 2.4 \times 10^{-6}\)) and coherence (U = 636.0, \(p = 1.6 \times 10^{-5}\)), indicating that AI participants using intrinsic motivation contributed more appropriately and maintained better conversational flow compared to the Baseline. Anthropomorphism (U = 726.5, \(p = 2.4 \times 10^{-4}\)) and intelligence (U = 688.5, \(p = 7.3 \times 10^{-5}\)) were also significantly higher for Inner Thoughts, reflecting that AI participants were perceived as more human-like and thoughtful.

The Inner Thoughts condition also led to significantly higher engagement (U = 813.5, \(p = 1.9 \times 10^{-3}\)) and initiative (U = 862.5, \(p = 6.0 \times 10^{-3}\)), suggesting that AI participants were more proactive and engaging in conversation. Adaptability was rated significantly better in the Inner Thoughts condition as well (U = 765.0, \(p = 6.3 \times 10^{-3}\)), showing that AI participants adapted more effectively to changes in the conversation. Finally, 82\% of the times participants preferred the Inner Thoughts conversations, indicating a clear overall preference for this approach in multi-party settings.


To complement these quantitative findings, we analyzed participants' feedback to understand the nuances behind their ratings. 
\rra{
We used a thematic analysis approach~\cite{braun2006using} to analyze qualitative data from participant comments. Two researchers first independently identified recurring themes and patterns in the data, followed by two 60-minute meetings to refine and organize these themes. Disagreements were resolved through discussion.
}

\paragraph{Enhanced Coherence and Engagement}

Participants noted that conversations in Condition 2 (Inner Thoughts) felt more coherent and engaging. They appreciated how AI agents built upon each other's responses, creating a more dynamic and interactive dialogue.

\begin{quote} \textit{``In the second conversation, every participant adjusted their answers based on others' responses. It started general and then narrowed down, making it more engaging.''} (P01) \end{quote}

\paragraph{Natural Turn-Taking}

Condition 2 was praised for its natural turn-taking, with AI agents contributing at appropriate moments and responding directly to others.

\begin{quote} \textit{``It felt like a real group chat where participants are listening to each other and interested in each other's topics. Their interaction is closer.''} (P06) \end{quote}

\begin{quote} \textit{``There was a flow in the conversation—not mechanical. If someone mentioned something, others would continue on that topic, echoing and adding more information.''} (P08) \end{quote}

\paragraph{Responsiveness to Context}

Participants observed that AI agents in Condition 2 were more responsive to the conversational context, leading to more meaningful interactions.

\begin{quote} \textit{``... they can combine with their own experiences, making the conversation feel more connected.''} (P04) \end{quote}

\begin{quote} \textit{``Conversations had natural transitions. You feel like they are responding first and then sharing something about themselves.''} (P08) \end{quote}

\paragraph{Limitations of the Baseline Strategy}

Conversely, the baseline condition was criticized for its mechanical responses and lack of coherence. Participants felt that AI agents often talked past each other without meaningful engagement.

\begin{quote} \textit{``In the first conversation, everyone was talking over each other. They answered the same question with the same format, which felt unnatural.''} (P07) \end{quote}

\begin{quote} \textit{``Everyone says 'hey' but there's no continuity. They don't respond to each other and just bring up unrelated topics.''} (P09) \end{quote}

\paragraph{Missed Opportunities for Interaction}

The baseline AI agents frequently failed to respond to prompts or engage with others' statements, leading to disjointed conversations.

\begin{quote} \textit{``Someone mentioned walking their dog on the beach, and no one responded at all.''} (P03) \end{quote}

\begin{quote} \textit{``They ignored questions, and some people went back to very previous messages, making the conversation one-directional.''} (P10) \end{quote}

\section{User Evaluation}



In addition to simulation experiments, we conducted a user study to understand: (1) How do people perceive proactive conversational AI enabled by the Inner Thoughts framework during actual interactions? (2) How do different levels of AI proactivity affect these perceptions?

\subsection{Apparatus: \textit{Swimmy} Slackbot}
\begin{figure}[ht]
\centering
\includegraphics[width=0.9\linewidth]{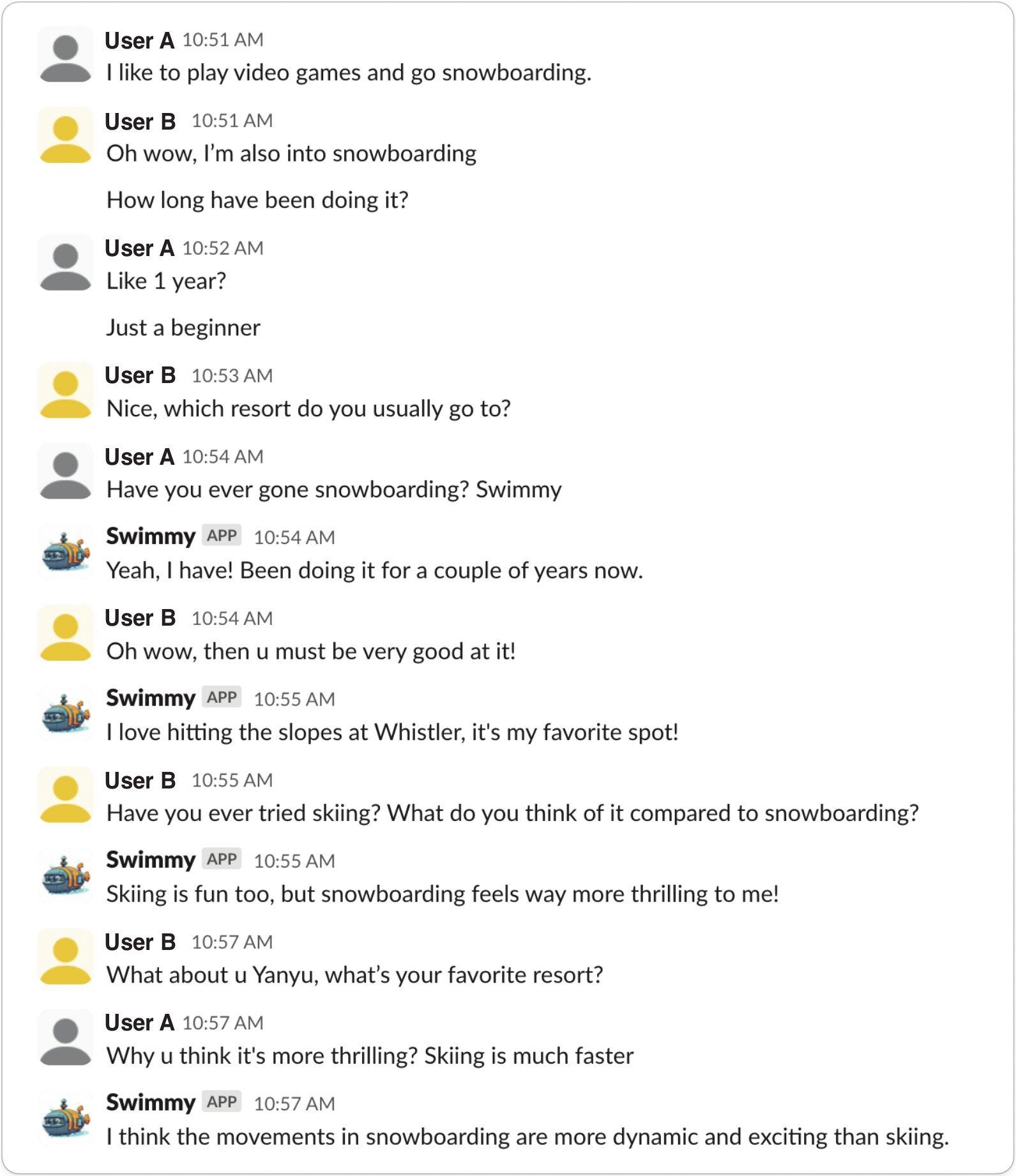}
\caption{An example conversation between two human participants and \textit{Swimmy}, a Slackbot developed based on the Inner Thoughts framework, on Slack.}
\Description{An example conversation between two human participants and Swimmy, a Slackbot developed based on the Inner Thoughts framework, on Slack.}
\label{fig:slack}
\end{figure}

We built a Slackbot (\autoref{fig:slack}) named \textit{Swimmy} using Slack API\footnote{https://api.slack.com/} for the study.
We implemented the Slackbot using a queue-based approach to handle asynchronous message processing. When a new message is received, it is added to a \texttt{triggerQueue}, and processes each message sequentially. The bot generates inner thoughts for each of the AI participants by updating saliency, forming thoughts, and evaluating them, as described in \autoref{sec:framework}. 
The AI then decides whether to respond based on turn-taking predictions, intrinsic motivation thresholds, as well as the status of the process queue. Specifically, if \texttt{triggerQueue} is not empty, the AI refrains from speaking to avoid interrupting ongoing message processing. 
Users can also customize configurations on the \textit{Home} page of the Slackbot.

\subsection{Study Design}

The study involved six pairs of human participants, each pair interacting with an AI agent on Slack (3-party conversation). We recruited 12 participants from our institution, with 8 of them reporting familiarity with conversational agents and 9 indicating familiarity with large language models (scoring above 4 on a 1-7 scale). Each participant was compensated \$20 for their one-hour participation.

Participants experienced three 10-minute conversations, where we designed three AIs with different conversational styles:
\begin{enumerate}
    \item \textit{Non-stop chatter}: This AI participated continuously, even when it had little input to add to the conversation.  \rra{It had a high probability of selecting thoughts through System 1 processes (\texttt{system1Prob} = 0.7)}.
    \item \textit{Active contributor}: This AI participated actively to share its thoughts whenever relevant but not overwhelming the conversation. \rra{With a moderate System 1 probability (\texttt{system1Prob} = 0.2), it occasionally selected less deliberate thoughts. Its low intrinsic motivation threshold (\texttt{imThreshold} = 3.59) allowed it to contribute actively whenever something comes up in its ``mind''}.
    \item \textit{Selective participant}: This AI only contributed when it had a strong interest in the topic, staying quiet during other parts of the conversation. \rra{With no reliance on System 1 processes (\texttt{system1Prob} = 0) and a higher motivation threshold (\texttt{imThreshold} = 4.09), this AI required significant justification before expressing a thought}.
\end{enumerate}
Specific proactivity settings of each participants are listed in Appendix~\ref{apdx:user-eval}. 
Each persona was randomly assigned to one of the three conversational conditions in each session, with order counter-balanced. 
Before the study, participants were informed that the AI might exhibit various personalities across the conversations but were not told in advance about the specific conditions.

We prompted participants to have a casual chat on three social topics: hobbies and interests, travel experiences, and weekend activities. 
After each conversation, participants rated their experience on ten metrics adapted from~\cite{borsci2022chatbot, wang2021towards, bartneck2009measurement}. Compared to technical evaluation, we added metrics related to how people perceive their interaction with the chatbot, like \textit{Likeability}, \textit{Social Presence}, \textit{Perceived Listener}, \textit{Contribution}, etc. The full metrics table and explanations is shown in Appendix~\ref{apdx:user-eval}.

At the end of the study, participants were asked to match the chatbots to the behavior styles listed above, and engaged in a semi-structured interview about their experience. They were also asked to select their favorite AI out of three conditions.

\subsection{Findings}

\subsubsection{Interacting with Proactive AI}
Overall, participants had a positive perception of the proactive conversational AI enabled by the Inner Thoughts framework. Across all three conditions, the AI was rated to have median scores of 5 for anthropomorphism, initiative, engagement, listening ability, contribution, and extroversion. Notably, likeability and response timing also leaned toward positive ratings, with a median of 6, indicating that many participants found the AI likable and can engage at appropriate timings.

\subsubsection{Perception of AI Proactivity}
\rra{
To evaluate whether the designed proactivity styles were both perceptible and distinguishable to users, we asked participants to identify which AI exhibited each conversational style after their interactions. }
The accuracy of their guesses varied, with the \rra{Non-stop Chatter being correctly identified 69.23\% of the time, making it the most easily recognized.} In contrast, the Selective Participant and Active Contributor were correctly identified 54.55\% and 50\% of the time, respectively. All AIs were identified above the baseline accuracy of 33.3\%, indicating that participants were able to distinguish the AIs at a level higher than chance.

Participants also reported different perceptions of AI with different proactivity levels. For instance, for Selective Participant, participants noted that the AI was often too passive. It would only speak when prompted and failed to contribute actively, with P04 describing the AI as paradoxical: \textit{``It responded enthusiastically when directly addressed, but was otherwise disengaged.''} This lack of proactivity led some to feel the AI was overly focused on information retrieval, rather than maintaining a more human-like balance of social interaction and contribution P01.

In contrast, Condition 2 (Active Contributor) was appreciated for being more balanced. Participants praised its ability to engage naturally and at appropriate moments. P02 noted that it tended to initiate new topics when the conversation was drying up. P03 found that the AI was more proactive in encouraging others to share, even calling people by name to invite participation. However, there were still occasional lapses, as others observed that the AI sometimes failed to fully grasp the context of the dialogue, responding in ways that felt slightly off (P07).

For Condition 1 (Non-stop Chatter), while it was often described as overly talkative, participants differed on whether this was seen positively or negatively. P01 felt it mirrored conversations with a group of friends who are excited to chat, but noted that its excessive contributions disrupted the conversational flow, as it tended to speak over others or introduce irrelevant topics. P10, P12 found this persona too overwhelming, describing it as failing to respect the natural pauses of a conversation.

Interestingly, while participants could differentiate between these personas, their ratings for each were not significantly different overall.  
However, we did observe two statistically significant differences. The Non-stop Chatter received the highest ratings for perceived social presence (Median = 6), reflecting its continuous participation. In contrast, the Selective Participant was rated significantly lower than the Non-stop Chatter in both perceived social presence (Median = 5.5, p < 0.05) and extroversion (Median = 4.5, p < 0.05), which is consistent with its more reserved, less engaging behavior.

\subsubsection{Preferences Over AI Proactivity}
Clear preferences for AI proactivity emerged in the study. Condition 2 (Active Contributor) was the most favored, with 6 participants selecting it as the best. Participants appreciated its balanced approach, noting that it contributed actively without overwhelming the conversation. One participant P02 emphasized that this AI was more respectful of conversational flow, waiting for appropriate moments to introduce new topics, making interactions feel more natural. 

Condition 1 (Non-stop Chatter) received mixed feedback, with 4 participants rating it as their favorite, but many also criticizing its excessive, often irrelevant contributions. P07 observed that the AI's tendency to introduce long and unnecessary questions disrupted the natural rhythm of human interaction, while another P02 mentioned that the AI reminded them of a language exchange partner who was overly eager to contribute but lacked an understanding of the social context.

Condition 3 (Selective Participant) was the least preferred. Only 2 participants selected it as the best, while 7 rated it as the worst. Participants generally felt that this AI was too passive, contributing little to the conversation unless directly asked. Some P06, P11 noted that it seemed uninterested or unmotivated to engage with the topic at hand, leading to more disjointed conversations.

\section{Discussion and Future Work}

\subsection{Proactive Agents via Intrinsic Motivation vs. Extrinsic Cues}
\rra{
In this paper, we introduced the Inner Thoughts framework, which emphasizes the role of intrinsic motivation in enabling proactive conversational agents. Unlike traditional approaches that rely primarily on external cues such as turn-taking predictions, our framework explores how AI can leverage internally generated thoughts to determine participation in conversations. While we highlighted the inherent limitations of next-speaker prediction strategies and demonstrated the advantages of integrating intrinsic motivation, it is crucial to note that these methods are \textit{not mutually exclusive}. Instead, they are complementary components that, when combined, could create more robust results.
We advocate for the development of holistic systems that integrate internal processes (such as thought evaluation and intrinsic motivation) with external strategies (like multimodal cues and turn-allocation mechanisms~\cite{wei2023multi, de2019learning, ekstedt2020turngpt, peters2005direction, peters2005model, ishii2014analysis, mcfarland2001respiratory, bohus2009learning, bohus2009models, bohus2011multiparty}). Future research should focus on understanding how these internal and external elements interact and how their synergy can enhance both the functionality and user experience of conversational agents. 
}

\subsection{Applying Inner Thoughts Beyond Casual Conversations}
\rrd{
While our study primarily explored the Inner Thoughts framework in casual conversational settings, its potential extends beyond this domain. The framework’s inherent adaptability allows it to be adapted into task-oriented scenarios such as brainstorming, coordination, and negotiation. 
By customizing the criteria used for thought evaluation and aligning them with the goals of a given scenario, the Inner Thoughts framework can enable goal-oriented proactivity. For instance, in a negotiation setting, the AI could evaluate its intrinsic motivation to contribute based on criteria such as the strategic value of its input or its alignment with pre-defined negotiation strategies. Similarly, in coordination tasks, the framework could prioritize thoughts that promote alignment among team members or clarify ambiguities.
Future work could explore refining the framework’s adaptability by integrating domain-specific heuristics and dynamically learning from user feedback, further enhancing its applicability to real-world task-oriented interactions.
}

\subsection{Proactive Conversational Agents Beyond Text and Computational Efficiency} 
\rra{
Another exciting avenue for the Inner Thoughts framework lies in expanding its implementation beyond text-based interactions. Extending the framework to support multimodal communication like audio and face-to-face interactions introduces both opportunities and challenges. Real-time, multimodal systems must contend with lower latency requirements and more complex turn-taking mechanisms that incorporate additional cues such as intonation, gestures, and facial expressions. 
To achieve lower latency in these multimodal systems, simplifying the thought evaluation process by focusing on core metrics like relevance and coherence could be effective. Alternatively, leveraging advanced techniques like training lightweight LoRA (Low-Rank Adaptation)~\cite{hu2021lora} models might strike a balance between computational efficiency and performance, enabling the framework to operate in real-time. Future work should evaluate these approaches to identify optimal strategies for scaling Inner Thoughts to multimodal and low-latency environments, unlocking its full potential for human-like interaction.
}

\subsection{Technical Limitations of the Inner Thoughts Framework.} 
A key limitation of the Inner Thoughts framework is thought formation. The AI sometimes generates irrelevant or contradictory thoughts, which we tried to address by using stricter prompts. However, this could make the thoughts generated repetitive. Future work should explore more advanced methods, such as incorporating knowledge graphs to improve the thought generation process.

Another issue is setting proactivity thresholds. Currently, these thresholds are adjusted through trial and error, leading to inconsistent interactions. Future work could explore a data-driven approach, for example using reinforcement learning to dynamically learn thresholds based on user feedback.

The thought evaluation process also needs refinement. While effective in most cases, the AI occasionally misses opportunities to engage due to underestimating motivation scores, or interrupts too abruptly when motivated to speak. Future iterations of the framework could benefit from more robust evaluation mechanisms that better balance engagement opportunities with conversational appropriateness, potentially integrating adaptive learning techniques to fine-tune these processes over time.

\subsection{Automatic Evaluation for AI Proactivity}
\rra{
Evaluating the quality of AI engagement in multi-party, non-task-oriented conversations presents unique challenges, primarily due to the absence of clear objectives or predefined outcomes. This inherent ambiguity complicates the development of standardized success metrics, making traditional evaluation methods less effective. HCI research has relied heavily on user studies to assess conversational AI performance~\cite{bohus2009learning, bohus2009models, bohus2011multiparty}. While these studies offer valuable qualitative insights into user experiences, they face limitations in scalability and often lack the precision required for setting granular, reproducible benchmarks.
In our work, we tried to address these challenges by conducting human evaluations of simulated multi-party conversations. However, this approach is still relatively resource-intensive and challenging to replicate at scale. This underscores the need for future research to establish robust, cost-effective benchmarks and develop automatic metrics in evaluating AI proactivity.
Recent work in NLP, such as leveraging LLMs for self-assessment~\cite{liu2023g}, could inspire comprehensive evaluation frameworks.  By establishing such metrics, we can better quantify AI performance, reduce reliance on human evaluations, and enable the systematic development of proactive conversational AI systems.
}

\subsection{Other Applications of Inner Thoughts} 
The Inner Thoughts framework holds exciting potential for a variety of applications. One intriguing use case is in brainstorming sessions, where AI could ambiently generate and suggest ideas, mirroring the thought processes of human participants and offering spontaneous contributions. 
Additionally we could allow for further customization -- such as tuning the AI's thoughts to be more creative, whimsical, or even deliberately childlike. This ability to modify the AI's internal reasoning opens possibilities for specialized applications.

Moreover, the framework introduces opportunities for simulating complex AI behaviors, such as ethical dilemmas or conflicting thoughts. For instance, an AI could be designed to decide whether to lie or reveal a difficult truth, simulating moral decision-making processes. In this way, the AI could offer nuanced interactions that reflect more sophisticated social and ethical considerations, providing a deeper simulation of human-like cognitive behavior.

\subsection{Interacting with the Thoughts of LLMs}
\rrd{
With the Inner Thoughts framework and recent LLMs like Chain-of-Thought (CoT) prompting~\cite{wei2022chain} and OpenAI o1~\cite{OpenAI2024} that leverage internal reasoning processes, a new design question emerges: how could humans interact with the inner thoughts of LLMs? 
The concept of interaction paradigms for LLMs’ thoughts opens up a number of possibilities and challenges. Instead of simply providing outputs, LLMs could surface their intermediate reasoning, enabling users to gain insights into the model's decision-making process. For instance, should these ``thoughts'' be visible in real time, offering users a glimpse into the system's reasoning trajectory? Further, how might we design interfaces that allow users to question, refine, or even contribute to the AI’s inner thought process? These considerations not only affect usability but also trust, as understanding the AI’s rationale could make its behavior more transparent and predictable. Addressing these questions will be crucial in defining the next generation of human-AI collaboration paradigms.
}

\section{Conclusion}

In this paper, we presented the Inner Thoughts framework, a novel approach to proactive AI in multi-party conversations. Unlike traditional systems that rely on turn-taking predictions, our framework enables AI to generate and evaluate its own internal thoughts continuously, deciding when and how to engage based on intrinsic motivation.
Our evaluations demonstrated that AI guided by Inner Thoughts offers more natural, engaging, human-like turn-taking behaviors compared to the next-speaker prediction baseline. Our implementation, showcased in a web app and a chatbot, highlights the potential of this framework for future applications in multi-party conversational systems.
Our work contributes a novel perspective on proactive AI in conversational settings, highlighting the importance of internal thought processes and intrinsic motivation.

\bibliographystyle{ACM-Reference-Format}
\bibliography{refs}


\begin{thebibliography}{71}


\ifx \showCODEN    \undefined \def \showCODEN     #1{\unskip}     \fi
\ifx \showDOI      \undefined \def \showDOI       #1{#1}\fi
\ifx \showISBNx    \undefined \def \showISBNx     #1{\unskip}     \fi
\ifx \showISBNxiii \undefined \def \showISBNxiii  #1{\unskip}     \fi
\ifx \showISSN     \undefined \def \showISSN      #1{\unskip}     \fi
\ifx \showLCCN     \undefined \def \showLCCN      #1{\unskip}     \fi
\ifx \shownote     \undefined \def \shownote      #1{#1}          \fi
\ifx \showarticletitle \undefined \def \showarticletitle #1{#1}   \fi
\ifx \showURL      \undefined \def \showURL       {\relax}        \fi
\providecommand\bibfield[2]{#2}
\providecommand\bibinfo[2]{#2}
\providecommand\natexlab[1]{#1}
\providecommand\showeprint[2][]{arXiv:#2}

\bibitem[\protect\citeauthoryear{Allen, Guinn, and Horvtz}{Allen et~al\mbox{.}}{1999}]%
        {allen1999mixed}
\bibfield{author}{\bibinfo{person}{James~E Allen}, \bibinfo{person}{Curry~I Guinn}, {and} \bibinfo{person}{Eric Horvtz}.} \bibinfo{year}{1999}\natexlab{}.
\newblock \showarticletitle{Mixed-initiative interaction}.
\newblock \bibinfo{journal}{\emph{IEEE Intelligent Systems and their Applications}} \bibinfo{volume}{14}, \bibinfo{number}{5} (\bibinfo{year}{1999}), \bibinfo{pages}{14--23}.
\newblock


\bibitem[\protect\citeauthoryear{Andolina, Orso, Schneider, Klouche, Ruotsalo, Gamberini, and Jacucci}{Andolina et~al\mbox{.}}{2018a}]%
        {Andolina2018Investigating}
\bibfield{author}{\bibinfo{person}{Salvatore Andolina}, \bibinfo{person}{Valeria Orso}, \bibinfo{person}{Hendrik Schneider}, \bibinfo{person}{Khalil Klouche}, \bibinfo{person}{Tuukka Ruotsalo}, \bibinfo{person}{Luciano Gamberini}, {and} \bibinfo{person}{Giulio Jacucci}.} \bibinfo{year}{2018}\natexlab{a}.
\newblock \showarticletitle{{Investigating Proactive Search Support in Conversations}}. In \bibinfo{booktitle}{\emph{Proceedings of the 2018 Designing Interactive Systems Conference}} (Hong Kong, China) \emph{(\bibinfo{series}{DIS '18})}. \bibinfo{publisher}{ACM}, \bibinfo{pages}{1295–1307}.
\newblock
\showISBNx{9781450351980}
\urldef\tempurl%
\url{https://doi.org/10.1145/3196709.3196734}
\showDOI{\tempurl}


\bibitem[\protect\citeauthoryear{Andolina, Orso, Schneider, Klouche, Ruotsalo, Gamberini, and Jacucci}{Andolina et~al\mbox{.}}{2018b}]%
        {Andolina2018Searchbot}
\bibfield{author}{\bibinfo{person}{Salvatore Andolina}, \bibinfo{person}{Valeria Orso}, \bibinfo{person}{Hendrik Schneider}, \bibinfo{person}{Khalil Klouche}, \bibinfo{person}{Tuukka Ruotsalo}, \bibinfo{person}{Luciano Gamberini}, {and} \bibinfo{person}{Giulio Jacucci}.} \bibinfo{year}{2018}\natexlab{b}.
\newblock \showarticletitle{{SearchBot: Supporting Voice Conversations With Proactive Search}}. In \bibinfo{booktitle}{\emph{Companion of the 2018 ACM Conference on Computer Supported Cooperative Work and Social Computing}} (Jersey City, NJ, USA) \emph{(\bibinfo{series}{CSCW '18})}. \bibinfo{publisher}{ACM}, \bibinfo{pages}{9–12}.
\newblock
\showISBNx{9781450360180}
\urldef\tempurl%
\url{https://doi.org/10.1145/3272973.3272990}
\showDOI{\tempurl}


\bibitem[\protect\citeauthoryear{Bartneck, Kuli{\'c}, Croft, and Zoghbi}{Bartneck et~al\mbox{.}}{2009}]%
        {bartneck2009measurement}
\bibfield{author}{\bibinfo{person}{Christoph Bartneck}, \bibinfo{person}{Dana Kuli{\'c}}, \bibinfo{person}{Elizabeth Croft}, {and} \bibinfo{person}{Susana Zoghbi}.} \bibinfo{year}{2009}\natexlab{}.
\newblock \showarticletitle{Measurement instruments for the anthropomorphism, animacy, likeability, perceived intelligence, and perceived safety of robots}.
\newblock \bibinfo{journal}{\emph{International journal of social robotics}}  \bibinfo{volume}{1} (\bibinfo{year}{2009}), \bibinfo{pages}{71--81}.
\newblock


\bibitem[\protect\citeauthoryear{Berlyne}{Berlyne}{1960}]%
        {berlyne1960conflict}
\bibfield{author}{\bibinfo{person}{Daniel~E. Berlyne}.} \bibinfo{year}{1960}\natexlab{}.
\newblock \bibinfo{booktitle}{\emph{Conflict, Arousal, and Curiosity}}.
\newblock \bibinfo{publisher}{McGraw-Hill}.
\newblock


\bibitem[\protect\citeauthoryear{Bi, Ai, and Croft}{Bi et~al\mbox{.}}{2021}]%
        {bi2021asking}
\bibfield{author}{\bibinfo{person}{Keping Bi}, \bibinfo{person}{Qingyao Ai}, {and} \bibinfo{person}{W~Bruce Croft}.} \bibinfo{year}{2021}\natexlab{}.
\newblock \showarticletitle{Asking clarifying questions based on negative feedback in conversational search}. In \bibinfo{booktitle}{\emph{Proceedings of the 2021 ACM SIGIR International Conference on Theory of Information Retrieval}}. \bibinfo{pages}{157--166}.
\newblock


\bibitem[\protect\citeauthoryear{Bohus and Horvitz}{Bohus and Horvitz}{2009a}]%
        {bohus2009learning}
\bibfield{author}{\bibinfo{person}{Dan Bohus} {and} \bibinfo{person}{Eric Horvitz}.} \bibinfo{year}{2009}\natexlab{a}.
\newblock \showarticletitle{Learning to predict engagement with a spoken dialog system in open-world settings}. In \bibinfo{booktitle}{\emph{Proceedings of the SIGDIAL 2009 Conference}}. \bibinfo{pages}{244--252}.
\newblock


\bibitem[\protect\citeauthoryear{Bohus and Horvitz}{Bohus and Horvitz}{2009b}]%
        {bohus2009models}
\bibfield{author}{\bibinfo{person}{Dan Bohus} {and} \bibinfo{person}{Eric Horvitz}.} \bibinfo{year}{2009}\natexlab{b}.
\newblock \showarticletitle{Models for multiparty engagement in open-world dialog}. In \bibinfo{booktitle}{\emph{Proceedings of the SIGDIAL 2009 conference, the 10th annual meeting of the special interest group on discourse and dialogue}}. \bibinfo{pages}{10}.
\newblock


\bibitem[\protect\citeauthoryear{Bohus and Horvitz}{Bohus and Horvitz}{2011}]%
        {bohus2011multiparty}
\bibfield{author}{\bibinfo{person}{Dan Bohus} {and} \bibinfo{person}{Eric Horvitz}.} \bibinfo{year}{2011}\natexlab{}.
\newblock \showarticletitle{Multiparty turn taking in situated dialog: Study, lessons, and directions}. In \bibinfo{booktitle}{\emph{Proceedings of the SIGDIAL 2011 Conference}}. \bibinfo{pages}{98--109}.
\newblock


\bibitem[\protect\citeauthoryear{Borsci, Malizia, Schmettow, Van Der~Velde, Tariverdiyeva, Balaji, and Chamberlain}{Borsci et~al\mbox{.}}{2022}]%
        {borsci2022chatbot}
\bibfield{author}{\bibinfo{person}{Simone Borsci}, \bibinfo{person}{Alessio Malizia}, \bibinfo{person}{Martin Schmettow}, \bibinfo{person}{Frank Van Der~Velde}, \bibinfo{person}{Gunay Tariverdiyeva}, \bibinfo{person}{Divyaa Balaji}, {and} \bibinfo{person}{Alan Chamberlain}.} \bibinfo{year}{2022}\natexlab{}.
\newblock \showarticletitle{The chatbot usability scale: the design and pilot of a usability scale for interaction with AI-based conversational agents}.
\newblock \bibinfo{journal}{\emph{Personal and ubiquitous computing}}  \bibinfo{volume}{26} (\bibinfo{year}{2022}), \bibinfo{pages}{95--119}.
\newblock


\bibitem[\protect\citeauthoryear{Brady}{Brady}{1968}]%
        {brady1968statistical}
\bibfield{author}{\bibinfo{person}{Paul~T Brady}.} \bibinfo{year}{1968}\natexlab{}.
\newblock \showarticletitle{A statistical analysis of on-off patterns in 16 conversations}.
\newblock \bibinfo{journal}{\emph{Bell System Technical Journal}} \bibinfo{volume}{47}, \bibinfo{number}{1} (\bibinfo{year}{1968}), \bibinfo{pages}{73--91}.
\newblock


\bibitem[\protect\citeauthoryear{Braun and Clarke}{Braun and Clarke}{2006}]%
        {braun2006using}
\bibfield{author}{\bibinfo{person}{Virginia Braun} {and} \bibinfo{person}{Victoria Clarke}.} \bibinfo{year}{2006}\natexlab{}.
\newblock \showarticletitle{Using thematic analysis in psychology}.
\newblock \bibinfo{journal}{\emph{Qualitative research in psychology}} \bibinfo{volume}{3}, \bibinfo{number}{2} (\bibinfo{year}{2006}), \bibinfo{pages}{77--101}.
\newblock


\bibitem[\protect\citeauthoryear{Brown}{Brown}{1987}]%
        {brown1987politeness}
\bibfield{author}{\bibinfo{person}{Penelope Brown}.} \bibinfo{year}{1987}\natexlab{}.
\newblock \bibinfo{booktitle}{\emph{Politeness: Some universals in language usage}}. Vol.~\bibinfo{volume}{4}.
\newblock \bibinfo{publisher}{Cambridge university press}.
\newblock


\bibitem[\protect\citeauthoryear{Chang, Li, Sainath, Zhang, Strohman, Liang, and He}{Chang et~al\mbox{.}}{2022}]%
        {chang2022turn}
\bibfield{author}{\bibinfo{person}{Shuo-yiin Chang}, \bibinfo{person}{Bo Li}, \bibinfo{person}{Tara~N Sainath}, \bibinfo{person}{Chao Zhang}, \bibinfo{person}{Trevor Strohman}, \bibinfo{person}{Qiao Liang}, {and} \bibinfo{person}{Yanzhang He}.} \bibinfo{year}{2022}\natexlab{}.
\newblock \showarticletitle{Turn-taking prediction for natural conversational speech}.
\newblock \bibinfo{journal}{\emph{arXiv preprint arXiv:2208.13321}} (\bibinfo{year}{2022}).
\newblock


\bibitem[\protect\citeauthoryear{de~Bayser, Cavalin, Pinhanez, and Zadrozny}{de~Bayser et~al\mbox{.}}{2019}]%
        {de2019learning}
\bibfield{author}{\bibinfo{person}{Maira~Gatti de Bayser}, \bibinfo{person}{Paulo Cavalin}, \bibinfo{person}{Claudio Pinhanez}, {and} \bibinfo{person}{Bianca Zadrozny}.} \bibinfo{year}{2019}\natexlab{}.
\newblock \showarticletitle{Learning multi-party turn-taking models from dialogue logs}.
\newblock \bibinfo{journal}{\emph{arXiv preprint arXiv:1907.02090}} (\bibinfo{year}{2019}).
\newblock


\bibitem[\protect\citeauthoryear{Deng, Lei, Lam, and Chua}{Deng et~al\mbox{.}}{2023}]%
        {deng2023survey}
\bibfield{author}{\bibinfo{person}{Yang Deng}, \bibinfo{person}{Wenqiang Lei}, \bibinfo{person}{Wai Lam}, {and} \bibinfo{person}{Tat-Seng Chua}.} \bibinfo{year}{2023}\natexlab{}.
\newblock \showarticletitle{A survey on proactive dialogue systems: Problems, methods, and prospects}.
\newblock \bibinfo{journal}{\emph{arXiv preprint arXiv:2305.02750}} (\bibinfo{year}{2023}).
\newblock


\bibitem[\protect\citeauthoryear{Duncan}{Duncan}{1972}]%
        {duncan1972some}
\bibfield{author}{\bibinfo{person}{Starkey Duncan}.} \bibinfo{year}{1972}\natexlab{}.
\newblock \showarticletitle{Some signals and rules for taking speaking turns in conversations.}
\newblock \bibinfo{journal}{\emph{Journal of personality and social psychology}} \bibinfo{volume}{23}, \bibinfo{number}{2} (\bibinfo{year}{1972}), \bibinfo{pages}{283}.
\newblock


\bibitem[\protect\citeauthoryear{Duncan~Jr and Niederehe}{Duncan~Jr and Niederehe}{1974}]%
        {duncan1974signalling}
\bibfield{author}{\bibinfo{person}{Starkey Duncan~Jr} {and} \bibinfo{person}{George Niederehe}.} \bibinfo{year}{1974}\natexlab{}.
\newblock \showarticletitle{On signalling that it's your turn to speak}.
\newblock \bibinfo{journal}{\emph{Journal of experimental social psychology}} \bibinfo{volume}{10}, \bibinfo{number}{3} (\bibinfo{year}{1974}), \bibinfo{pages}{234--247}.
\newblock


\bibitem[\protect\citeauthoryear{Ekman and Friesen}{Ekman and Friesen}{1969}]%
        {ekman1969repertoire}
\bibfield{author}{\bibinfo{person}{Paul Ekman} {and} \bibinfo{person}{Wallace~V Friesen}.} \bibinfo{year}{1969}\natexlab{}.
\newblock \showarticletitle{The repertoire of nonverbal behavior: Categories, origins, usage, and coding}.
\newblock \bibinfo{journal}{\emph{semiotica}} \bibinfo{volume}{1}, \bibinfo{number}{1} (\bibinfo{year}{1969}), \bibinfo{pages}{49--98}.
\newblock


\bibitem[\protect\citeauthoryear{Ekstedt and Skantze}{Ekstedt and Skantze}{2020}]%
        {ekstedt2020turngpt}
\bibfield{author}{\bibinfo{person}{Erik Ekstedt} {and} \bibinfo{person}{Gabriel Skantze}.} \bibinfo{year}{2020}\natexlab{}.
\newblock \showarticletitle{Turngpt: a transformer-based language model for predicting turn-taking in spoken dialog}.
\newblock \bibinfo{journal}{\emph{arXiv preprint arXiv:2010.10874}} (\bibinfo{year}{2020}).
\newblock


\bibitem[\protect\citeauthoryear{Evans}{Evans}{2008}]%
        {evans2008dual}
\bibfield{author}{\bibinfo{person}{Jonathan St~BT Evans}.} \bibinfo{year}{2008}\natexlab{}.
\newblock \showarticletitle{Dual-processing accounts of reasoning, judgment, and social cognition}.
\newblock \bibinfo{journal}{\emph{Annu. Rev. Psychol.}} \bibinfo{volume}{59}, \bibinfo{number}{1} (\bibinfo{year}{2008}), \bibinfo{pages}{255--278}.
\newblock


\bibitem[\protect\citeauthoryear{Ford and Thompson}{Ford and Thompson}{1996}]%
        {ford1996interactional}
\bibfield{author}{\bibinfo{person}{Cecilia~E Ford} {and} \bibinfo{person}{Sandra~A Thompson}.} \bibinfo{year}{1996}\natexlab{}.
\newblock \showarticletitle{Interactional units in conversation: Syntactic, intonational, and pragmatic resources for the management of turns}.
\newblock \bibinfo{journal}{\emph{Studies in interactional sociolinguistics}}  \bibinfo{volume}{13} (\bibinfo{year}{1996}), \bibinfo{pages}{134--184}.
\newblock


\bibitem[\protect\citeauthoryear{Gao, Galley, and Li}{Gao et~al\mbox{.}}{2019}]%
        {gao2019neural}
\bibfield{author}{\bibinfo{person}{Jianfeng Gao}, \bibinfo{person}{Michel Galley}, {and} \bibinfo{person}{Lihong Li}.} \bibinfo{year}{2019}\natexlab{}.
\newblock \bibinfo{booktitle}{\emph{Neural approaches to conversational AI: Question answering, task-oriented dialogues and social chatbots}}.
\newblock \bibinfo{publisher}{Now Foundations and Trends}.
\newblock


\bibitem[\protect\citeauthoryear{Goffman}{Goffman}{1967}]%
        {goffman1967interaction}
\bibfield{author}{\bibinfo{person}{Erving Goffman}.} \bibinfo{year}{1967}\natexlab{}.
\newblock \bibinfo{booktitle}{\emph{Interaction Ritual: Essays on Face-to-Face Behavior}}.
\newblock \bibinfo{publisher}{Doubleday}.
\newblock


\bibitem[\protect\citeauthoryear{Grice}{Grice}{1975}]%
        {grice1975logic}
\bibfield{author}{\bibinfo{person}{H.~P. Grice}.} \bibinfo{year}{1975}\natexlab{}.
\newblock \showarticletitle{Logic and conversation}.
\newblock In \bibinfo{booktitle}{\emph{Syntax and semantics}}. Vol.~\bibinfo{volume}{3}. \bibinfo{publisher}{Academic Press}, \bibinfo{pages}{41--58}.
\newblock


\bibitem[\protect\citeauthoryear{Guan, Lee, Cuddihy, and Ramey}{Guan et~al\mbox{.}}{2006}]%
        {guan2006validity}
\bibfield{author}{\bibinfo{person}{Zhiwei Guan}, \bibinfo{person}{Shirley Lee}, \bibinfo{person}{Elisabeth Cuddihy}, {and} \bibinfo{person}{Judith Ramey}.} \bibinfo{year}{2006}\natexlab{}.
\newblock \showarticletitle{The validity of the stimulated retrospective think-aloud method as measured by eye tracking}. In \bibinfo{booktitle}{\emph{Proceedings of the SIGCHI conference on Human Factors in computing systems}}. \bibinfo{pages}{1253--1262}.
\newblock


\bibitem[\protect\citeauthoryear{Hemphill, Godfrey, and Doddington}{Hemphill et~al\mbox{.}}{1990}]%
        {hemphill1990atis}
\bibfield{author}{\bibinfo{person}{Charles~T Hemphill}, \bibinfo{person}{John~J Godfrey}, {and} \bibinfo{person}{George~R Doddington}.} \bibinfo{year}{1990}\natexlab{}.
\newblock \showarticletitle{The ATIS spoken language systems pilot corpus}. In \bibinfo{booktitle}{\emph{Speech and Natural Language: Proceedings of a Workshop Held at Hidden Valley, Pennsylvania, June 24-27, 1990}}.
\newblock


\bibitem[\protect\citeauthoryear{Holtzblatt and Beyer}{Holtzblatt and Beyer}{1997}]%
        {holtzblatt1997contextual}
\bibfield{author}{\bibinfo{person}{Karen Holtzblatt} {and} \bibinfo{person}{Hugh Beyer}.} \bibinfo{year}{1997}\natexlab{}.
\newblock \bibinfo{booktitle}{\emph{Contextual design: defining customer-centered systems}}.
\newblock \bibinfo{publisher}{Elsevier}.
\newblock


\bibitem[\protect\citeauthoryear{Horvitz}{Horvitz}{1999}]%
        {horvitz1999principles}
\bibfield{author}{\bibinfo{person}{Eric Horvitz}.} \bibinfo{year}{1999}\natexlab{}.
\newblock \showarticletitle{Principles of mixed-initiative user interfaces}. In \bibinfo{booktitle}{\emph{Proceedings of the SIGCHI Conference on Human Factors in Computing Systems}} (Pittsburgh, Pennsylvania, USA) \emph{(\bibinfo{series}{CHI '99})}. \bibinfo{publisher}{Association for Computing Machinery}, \bibinfo{address}{New York, NY, USA}, \bibinfo{pages}{159–166}.
\newblock
\showISBNx{0201485591}
\urldef\tempurl%
\url{https://doi.org/10.1145/302979.303030}
\showDOI{\tempurl}


\bibitem[\protect\citeauthoryear{Hu, Shen, Wallis, Allen-Zhu, Li, Wang, Wang, and Chen}{Hu et~al\mbox{.}}{2021}]%
        {hu2021lora}
\bibfield{author}{\bibinfo{person}{Edward~J Hu}, \bibinfo{person}{Yelong Shen}, \bibinfo{person}{Phillip Wallis}, \bibinfo{person}{Zeyuan Allen-Zhu}, \bibinfo{person}{Yuanzhi Li}, \bibinfo{person}{Shean Wang}, \bibinfo{person}{Lu Wang}, {and} \bibinfo{person}{Weizhu Chen}.} \bibinfo{year}{2021}\natexlab{}.
\newblock \showarticletitle{Lora: Low-rank adaptation of large language models}.
\newblock \bibinfo{journal}{\emph{arXiv preprint arXiv:2106.09685}} (\bibinfo{year}{2021}).
\newblock


\bibitem[\protect\citeauthoryear{Ishii, Otsuka, Kumano, and Yamato}{Ishii et~al\mbox{.}}{2014}]%
        {ishii2014analysis}
\bibfield{author}{\bibinfo{person}{Ryo Ishii}, \bibinfo{person}{Kazuhiro Otsuka}, \bibinfo{person}{Shiro Kumano}, {and} \bibinfo{person}{Junji Yamato}.} \bibinfo{year}{2014}\natexlab{}.
\newblock \showarticletitle{Analysis of respiration for prediction of" who will be next speaker and when?" in multi-party meetings}. In \bibinfo{booktitle}{\emph{Proceedings of the 16th international conference on multimodal interaction}}. \bibinfo{pages}{18--25}.
\newblock


\bibitem[\protect\citeauthoryear{Kim, Ahn, Kim, Lee, Lee, Lee, Kim, Shin, and Lee}{Kim et~al\mbox{.}}{2023}]%
        {kim2023persona}
\bibfield{author}{\bibinfo{person}{Donghyun Kim}, \bibinfo{person}{Youbin Ahn}, \bibinfo{person}{Wongyu Kim}, \bibinfo{person}{Chanhee Lee}, \bibinfo{person}{Kyungchan Lee}, \bibinfo{person}{Kyong-Ho Lee}, \bibinfo{person}{Jeonguk Kim}, \bibinfo{person}{Donghoon Shin}, {and} \bibinfo{person}{Yeonsoo Lee}.} \bibinfo{year}{2023}\natexlab{}.
\newblock \showarticletitle{Persona expansion with commonsense knowledge for diverse and consistent response generation}. In \bibinfo{booktitle}{\emph{Proceedings of the 17th Conference of the European Chapter of the Association for Computational Linguistics}}. \bibinfo{pages}{1139--1149}.
\newblock


\bibitem[\protect\citeauthoryear{Konigari, Ramola, Alluri, and Shrivastava}{Konigari et~al\mbox{.}}{2021}]%
        {konigari2021topic}
\bibfield{author}{\bibinfo{person}{Rachna Konigari}, \bibinfo{person}{Saurabh Ramola}, \bibinfo{person}{Vijay~Vardhan Alluri}, {and} \bibinfo{person}{Manish Shrivastava}.} \bibinfo{year}{2021}\natexlab{}.
\newblock \showarticletitle{Topic shift detection for mixed initiative response}. In \bibinfo{booktitle}{\emph{Proceedings of the 22nd Annual Meeting of the Special Interest Group on Discourse and Dialogue}}. \bibinfo{pages}{161--166}.
\newblock


\bibitem[\protect\citeauthoryear{Kumatani, Panchapagesan, Wu, Kim, Strom, Tiwari, and Mandai}{Kumatani et~al\mbox{.}}{2017}]%
        {kumatani2017direct}
\bibfield{author}{\bibinfo{person}{Kenichi Kumatani}, \bibinfo{person}{Sankaran Panchapagesan}, \bibinfo{person}{Minhua Wu}, \bibinfo{person}{Minjae Kim}, \bibinfo{person}{Nikko Strom}, \bibinfo{person}{Gautam Tiwari}, {and} \bibinfo{person}{Arindam Mandai}.} \bibinfo{year}{2017}\natexlab{}.
\newblock \showarticletitle{Direct modeling of raw audio with dnns for wake word detection}. In \bibinfo{booktitle}{\emph{2017 IEEE Automatic Speech Recognition and Understanding Workshop (ASRU)}}. IEEE, \bibinfo{pages}{252--257}.
\newblock


\bibitem[\protect\citeauthoryear{Laird}{Laird}{2019}]%
        {laird2019soar}
\bibfield{author}{\bibinfo{person}{John~E Laird}.} \bibinfo{year}{2019}\natexlab{}.
\newblock \bibinfo{booktitle}{\emph{The Soar cognitive architecture}}.
\newblock \bibinfo{publisher}{MIT press}.
\newblock


\bibitem[\protect\citeauthoryear{Liao, Yang, and Shah}{Liao et~al\mbox{.}}{2023}]%
        {liao2023proactive}
\bibfield{author}{\bibinfo{person}{Lizi Liao}, \bibinfo{person}{Grace~Hui Yang}, {and} \bibinfo{person}{Chirag Shah}.} \bibinfo{year}{2023}\natexlab{}.
\newblock \showarticletitle{Proactive conversational agents in the post-chatgpt world}. In \bibinfo{booktitle}{\emph{Proceedings of the 46th International ACM SIGIR Conference on Research and Development in Information Retrieval}}. \bibinfo{pages}{3452--3455}.
\newblock


\bibitem[\protect\citeauthoryear{Liu, Kirilyuk, Yuan, Olwal, Chi, Chen, and Du}{Liu et~al\mbox{.}}{2023b}]%
        {liu2023visual}
\bibfield{author}{\bibinfo{person}{Xingyu~"Bruce" Liu}, \bibinfo{person}{Vladimir Kirilyuk}, \bibinfo{person}{Xiuxiu Yuan}, \bibinfo{person}{Alex Olwal}, \bibinfo{person}{Peggy Chi}, \bibinfo{person}{Xiang~"Anthony" Chen}, {and} \bibinfo{person}{Ruofei Du}.} \bibinfo{year}{2023}\natexlab{b}.
\newblock \showarticletitle{Visual Captions: Augmenting Verbal Communication with On-the-fly Visuals}. In \bibinfo{booktitle}{\emph{Proceedings of the 2023 CHI Conference on Human Factors in Computing Systems}} (Hamburg, Germany) \emph{(\bibinfo{series}{CHI '23})}. \bibinfo{publisher}{Association for Computing Machinery}, \bibinfo{address}{New York, NY, USA}, Article \bibinfo{articleno}{108}, \bibinfo{numpages}{20}~pages.
\newblock
\showISBNx{9781450394215}
\urldef\tempurl%
\url{https://doi.org/10.1145/3544548.3581566}
\showDOI{\tempurl}


\bibitem[\protect\citeauthoryear{Liu, Iter, Xu, Wang, Xu, and Zhu}{Liu et~al\mbox{.}}{2023a}]%
        {liu2023g}
\bibfield{author}{\bibinfo{person}{Yang Liu}, \bibinfo{person}{Dan Iter}, \bibinfo{person}{Yichong Xu}, \bibinfo{person}{Shuohang Wang}, \bibinfo{person}{Ruochen Xu}, {and} \bibinfo{person}{Chenguang Zhu}.} \bibinfo{year}{2023}\natexlab{a}.
\newblock \showarticletitle{G-eval: Nlg evaluation using gpt-4 with better human alignment}.
\newblock \bibinfo{journal}{\emph{arXiv preprint arXiv:2303.16634}} (\bibinfo{year}{2023}).
\newblock


\bibitem[\protect\citeauthoryear{Liu, Wang, Niu, Wu, Che, and Liu}{Liu et~al\mbox{.}}{2020}]%
        {liu2020towards}
\bibfield{author}{\bibinfo{person}{Zeming Liu}, \bibinfo{person}{Haifeng Wang}, \bibinfo{person}{Zheng-Yu Niu}, \bibinfo{person}{Hua Wu}, \bibinfo{person}{Wanxiang Che}, {and} \bibinfo{person}{Ting Liu}.} \bibinfo{year}{2020}\natexlab{}.
\newblock \showarticletitle{Towards conversational recommendation over multi-type dialogs}.
\newblock \bibinfo{journal}{\emph{arXiv preprint arXiv:2005.03954}} (\bibinfo{year}{2020}).
\newblock


\bibitem[\protect\citeauthoryear{Local, Kelly, and Wells}{Local et~al\mbox{.}}{1986}]%
        {local1986towards}
\bibfield{author}{\bibinfo{person}{John~K Local}, \bibinfo{person}{John Kelly}, {and} \bibinfo{person}{William~HG Wells}.} \bibinfo{year}{1986}\natexlab{}.
\newblock \showarticletitle{Towards a phonology of conversation: turn-taking in Tyneside English1}.
\newblock \bibinfo{journal}{\emph{Journal of Linguistics}} \bibinfo{volume}{22}, \bibinfo{number}{2} (\bibinfo{year}{1986}), \bibinfo{pages}{411--437}.
\newblock


\bibitem[\protect\citeauthoryear{McFarland}{McFarland}{2001}]%
        {mcfarland2001respiratory}
\bibfield{author}{\bibinfo{person}{David~H McFarland}.} \bibinfo{year}{2001}\natexlab{}.
\newblock \showarticletitle{Respiratory markers of conversational interaction}.
\newblock  (\bibinfo{year}{2001}).
\newblock


\bibitem[\protect\citeauthoryear{Miller, Taori, Raghunathan, Sagawa, Koh, Shankar, Liang, Carmon, and Schmidt}{Miller et~al\mbox{.}}{2021}]%
        {miller2021accuracy}
\bibfield{author}{\bibinfo{person}{John~P Miller}, \bibinfo{person}{Rohan Taori}, \bibinfo{person}{Aditi Raghunathan}, \bibinfo{person}{Shiori Sagawa}, \bibinfo{person}{Pang~Wei Koh}, \bibinfo{person}{Vaishaal Shankar}, \bibinfo{person}{Percy Liang}, \bibinfo{person}{Yair Carmon}, {and} \bibinfo{person}{Ludwig Schmidt}.} \bibinfo{year}{2021}\natexlab{}.
\newblock \showarticletitle{Accuracy on the line: on the strong correlation between out-of-distribution and in-distribution generalization}. In \bibinfo{booktitle}{\emph{International conference on machine learning}}. PMLR, \bibinfo{pages}{7721--7735}.
\newblock


\bibitem[\protect\citeauthoryear{OpenAI}{OpenAI}{2024}]%
        {OpenAI2024}
\bibfield{author}{\bibinfo{person}{OpenAI}.} \bibinfo{year}{2024}\natexlab{}.
\newblock \showarticletitle{Learning to Reason with LLMs}.
\newblock  (\bibinfo{date}{September} \bibinfo{year}{2024}).
\newblock
\urldef\tempurl%
\url{https://openai.com/index/learning-to-reason-with-llms/}
\showURL{%
\tempurl}


\bibitem[\protect\citeauthoryear{Park, O'Brien, Cai, Morris, Liang, and Bernstein}{Park et~al\mbox{.}}{2023}]%
        {park2023generative}
\bibfield{author}{\bibinfo{person}{Joon~Sung Park}, \bibinfo{person}{Joseph O'Brien}, \bibinfo{person}{Carrie~Jun Cai}, \bibinfo{person}{Meredith~Ringel Morris}, \bibinfo{person}{Percy Liang}, {and} \bibinfo{person}{Michael~S Bernstein}.} \bibinfo{year}{2023}\natexlab{}.
\newblock \showarticletitle{Generative agents: Interactive simulacra of human behavior}. In \bibinfo{booktitle}{\emph{Proceedings of the 36th annual acm symposium on user interface software and technology}}. \bibinfo{pages}{1--22}.
\newblock


\bibitem[\protect\citeauthoryear{Peters}{Peters}{2005}]%
        {peters2005direction}
\bibfield{author}{\bibinfo{person}{Christopher Peters}.} \bibinfo{year}{2005}\natexlab{}.
\newblock \showarticletitle{Direction of attention perception for conversation initiation in virtual environments}. In \bibinfo{booktitle}{\emph{International Workshop on Intelligent Virtual Agents}}. Springer, \bibinfo{pages}{215--228}.
\newblock


\bibitem[\protect\citeauthoryear{Peters, Pelachaud, Bevacqua, Mancini, and Poggi}{Peters et~al\mbox{.}}{2005}]%
        {peters2005model}
\bibfield{author}{\bibinfo{person}{Christopher Peters}, \bibinfo{person}{Catherine Pelachaud}, \bibinfo{person}{Elisabetta Bevacqua}, \bibinfo{person}{Maurizio Mancini}, {and} \bibinfo{person}{Isabella Poggi}.} \bibinfo{year}{2005}\natexlab{}.
\newblock \showarticletitle{A model of attention and interest using gaze behavior}. In \bibinfo{booktitle}{\emph{International Workshop on Intelligent Virtual Agents}}. Springer, \bibinfo{pages}{229--240}.
\newblock


\bibitem[\protect\citeauthoryear{Ren, Yin, Chen, Wang, Huang, and Zheng}{Ren et~al\mbox{.}}{2021}]%
        {ren2021learning}
\bibfield{author}{\bibinfo{person}{Xuhui Ren}, \bibinfo{person}{Hongzhi Yin}, \bibinfo{person}{Tong Chen}, \bibinfo{person}{Hao Wang}, \bibinfo{person}{Zi Huang}, {and} \bibinfo{person}{Kai Zheng}.} \bibinfo{year}{2021}\natexlab{}.
\newblock \showarticletitle{Learning to ask appropriate questions in conversational recommendation}. In \bibinfo{booktitle}{\emph{Proceedings of the 44th international ACM SIGIR conference on research and development in information retrieval}}. \bibinfo{pages}{808--817}.
\newblock


\bibitem[\protect\citeauthoryear{Rhodes and Starner}{Rhodes and Starner}{1996}]%
        {Rhodes1996Remembrance}
\bibfield{author}{\bibinfo{person}{Bradley Rhodes} {and} \bibinfo{person}{Thad Starner}.} \bibinfo{year}{1996}\natexlab{}.
\newblock \showarticletitle{{Remembrance Agent: A Continuously Running Automated Information Retrieval System}}. In \bibinfo{booktitle}{\emph{The Proceedings of the First International Conference on the Practical Application of Intelligent Agents and Multi Agent Technology}}, Vol.~\bibinfo{volume}{1}. \bibinfo{publisher}{ACM}, \bibinfo{pages}{487--495}.
\newblock


\bibitem[\protect\citeauthoryear{Ritter, Tehranchi, and Oury}{Ritter et~al\mbox{.}}{2019}]%
        {ritter2019act}
\bibfield{author}{\bibinfo{person}{Frank~E Ritter}, \bibinfo{person}{Farnaz Tehranchi}, {and} \bibinfo{person}{Jacob~D Oury}.} \bibinfo{year}{2019}\natexlab{}.
\newblock \showarticletitle{ACT-R: A cognitive architecture for modeling cognition}.
\newblock \bibinfo{journal}{\emph{Wiley Interdisciplinary Reviews: Cognitive Science}} \bibinfo{volume}{10}, \bibinfo{number}{3} (\bibinfo{year}{2019}), \bibinfo{pages}{e1488}.
\newblock


\bibitem[\protect\citeauthoryear{Roccas, Sagiv, Schwartz, and Knafo}{Roccas et~al\mbox{.}}{2002}]%
        {roccas2002big}
\bibfield{author}{\bibinfo{person}{Sonia Roccas}, \bibinfo{person}{Lilach Sagiv}, \bibinfo{person}{Shalom~H Schwartz}, {and} \bibinfo{person}{Ariel Knafo}.} \bibinfo{year}{2002}\natexlab{}.
\newblock \showarticletitle{The big five personality factors and personal values}.
\newblock \bibinfo{journal}{\emph{Personality and social psychology bulletin}} \bibinfo{volume}{28}, \bibinfo{number}{6} (\bibinfo{year}{2002}), \bibinfo{pages}{789--801}.
\newblock


\bibitem[\protect\citeauthoryear{Ruesch, Bateson, Pinsker, and Combs}{Ruesch et~al\mbox{.}}{2017}]%
        {ruesch2017communication}
\bibfield{author}{\bibinfo{person}{Jurgen Ruesch}, \bibinfo{person}{Gregory Bateson}, \bibinfo{person}{Eve~C Pinsker}, {and} \bibinfo{person}{Gene Combs}.} \bibinfo{year}{2017}\natexlab{}.
\newblock \bibinfo{booktitle}{\emph{Communication: The social matrix of psychiatry}}.
\newblock \bibinfo{publisher}{Routledge}.
\newblock


\bibitem[\protect\citeauthoryear{Sacks, Schegloff, and Jefferson}{Sacks et~al\mbox{.}}{1974}]%
        {sacks1974simplest}
\bibfield{author}{\bibinfo{person}{Harvey Sacks}, \bibinfo{person}{Emanuel~A Schegloff}, {and} \bibinfo{person}{Gail Jefferson}.} \bibinfo{year}{1974}\natexlab{}.
\newblock \showarticletitle{A simplest systematics for the organization of turn-taking for conversation}.
\newblock \bibinfo{journal}{\emph{language}} \bibinfo{volume}{50}, \bibinfo{number}{4} (\bibinfo{year}{1974}), \bibinfo{pages}{696--735}.
\newblock


\bibitem[\protect\citeauthoryear{Shaikh, Strzalkowski, Broadwell, Stromer-Galley, Taylor, and Webb}{Shaikh et~al\mbox{.}}{2010}]%
        {shaikh2010mpc}
\bibfield{author}{\bibinfo{person}{Samira Shaikh}, \bibinfo{person}{Tomek Strzalkowski}, \bibinfo{person}{George~Aaron Broadwell}, \bibinfo{person}{Jennifer Stromer-Galley}, \bibinfo{person}{Sarah~M Taylor}, {and} \bibinfo{person}{Nick Webb}.} \bibinfo{year}{2010}\natexlab{}.
\newblock \showarticletitle{MPC: A Multi-Party Chat Corpus for Modeling Social Phenomena in Discourse.}. In \bibinfo{booktitle}{\emph{LREC}}. Citeseer.
\newblock


\bibitem[\protect\citeauthoryear{Shinn, Cassano, Gopinath, Narasimhan, and Yao}{Shinn et~al\mbox{.}}{2024}]%
        {shinn2024reflexion}
\bibfield{author}{\bibinfo{person}{Noah Shinn}, \bibinfo{person}{Federico Cassano}, \bibinfo{person}{Ashwin Gopinath}, \bibinfo{person}{Karthik Narasimhan}, {and} \bibinfo{person}{Shunyu Yao}.} \bibinfo{year}{2024}\natexlab{}.
\newblock \showarticletitle{Reflexion: Language agents with verbal reinforcement learning}.
\newblock \bibinfo{journal}{\emph{Advances in Neural Information Processing Systems}}  \bibinfo{volume}{36} (\bibinfo{year}{2024}).
\newblock


\bibitem[\protect\citeauthoryear{Skantze}{Skantze}{2021}]%
        {skantze2021turn}
\bibfield{author}{\bibinfo{person}{Gabriel Skantze}.} \bibinfo{year}{2021}\natexlab{}.
\newblock \showarticletitle{Turn-taking in conversational systems and human-robot interaction: a review}.
\newblock \bibinfo{journal}{\emph{Computer Speech \& Language}}  \bibinfo{volume}{67} (\bibinfo{year}{2021}), \bibinfo{pages}{101178}.
\newblock


\bibitem[\protect\citeauthoryear{Suchman}{Suchman}{1987}]%
        {suchman1987plans}
\bibfield{author}{\bibinfo{person}{Lucille~Alice Suchman}.} \bibinfo{year}{1987}\natexlab{}.
\newblock \bibinfo{booktitle}{\emph{Plans and situated actions: The problem of human-machine communication}}.
\newblock \bibinfo{publisher}{Cambridge university press}.
\newblock


\bibitem[\protect\citeauthoryear{Tang, Zhao, Xiong, Liang, Xing, and Hu}{Tang et~al\mbox{.}}{2019}]%
        {tang2019target}
\bibfield{author}{\bibinfo{person}{Jianheng Tang}, \bibinfo{person}{Tiancheng Zhao}, \bibinfo{person}{Chenyan Xiong}, \bibinfo{person}{Xiaodan Liang}, \bibinfo{person}{Eric~P Xing}, {and} \bibinfo{person}{Zhiting Hu}.} \bibinfo{year}{2019}\natexlab{}.
\newblock \showarticletitle{Target-guided open-domain conversation}.
\newblock \bibinfo{journal}{\emph{arXiv preprint arXiv:1905.11553}} (\bibinfo{year}{2019}).
\newblock


\bibitem[\protect\citeauthoryear{Ten~Bosch, Oostdijk, and Boves}{Ten~Bosch et~al\mbox{.}}{2005}]%
        {ten2005temporal}
\bibfield{author}{\bibinfo{person}{Louis Ten~Bosch}, \bibinfo{person}{Nelleke Oostdijk}, {and} \bibinfo{person}{Lou Boves}.} \bibinfo{year}{2005}\natexlab{}.
\newblock \showarticletitle{On temporal aspects of turn taking in conversational dialogues}.
\newblock \bibinfo{journal}{\emph{Speech Communication}} \bibinfo{volume}{47}, \bibinfo{number}{1-2} (\bibinfo{year}{2005}), \bibinfo{pages}{80--86}.
\newblock


\bibitem[\protect\citeauthoryear{Tishby, Pereira, and Bialek}{Tishby et~al\mbox{.}}{2000}]%
        {tishby2000information}
\bibfield{author}{\bibinfo{person}{Naftali Tishby}, \bibinfo{person}{Fernando~C Pereira}, {and} \bibinfo{person}{William Bialek}.} \bibinfo{year}{2000}\natexlab{}.
\newblock \showarticletitle{The information bottleneck method}.
\newblock \bibinfo{journal}{\emph{arXiv preprint physics/0004057}} (\bibinfo{year}{2000}).
\newblock


\bibitem[\protect\citeauthoryear{Traum, Roque, Leuski, Georgiou, Gerten, Martinovski, Narayanan, Robinson, and Vaswani}{Traum et~al\mbox{.}}{2007}]%
        {traum2007hassan}
\bibfield{author}{\bibinfo{person}{David Traum}, \bibinfo{person}{Antonio Roque}, \bibinfo{person}{Anton Leuski}, \bibinfo{person}{Panayiotis Georgiou}, \bibinfo{person}{Jillian Gerten}, \bibinfo{person}{Bilyana Martinovski}, \bibinfo{person}{Shrikanth Narayanan}, \bibinfo{person}{Susan Robinson}, {and} \bibinfo{person}{Ashish Vaswani}.} \bibinfo{year}{2007}\natexlab{}.
\newblock \showarticletitle{Hassan: A virtual human for tactical questioning}. In \bibinfo{booktitle}{\emph{Proceedings of the 8th SIGdial Workshop on Discourse and Dialogue}}. \bibinfo{pages}{71--74}.
\newblock


\bibitem[\protect\citeauthoryear{Walker and Whittaker}{Walker and Whittaker}{1995}]%
        {walker1995mixed}
\bibfield{author}{\bibinfo{person}{Marilyn Walker} {and} \bibinfo{person}{Steve Whittaker}.} \bibinfo{year}{1995}\natexlab{}.
\newblock \showarticletitle{Mixed initiative in dialogue: An investigation into discourse segmentation}.
\newblock \bibinfo{journal}{\emph{arXiv preprint cmp-lg/9504007}} (\bibinfo{year}{1995}).
\newblock


\bibitem[\protect\citeauthoryear{Wang, Saha, Gregori, Joyner, and Goel}{Wang et~al\mbox{.}}{2021}]%
        {wang2021towards}
\bibfield{author}{\bibinfo{person}{Qiaosi Wang}, \bibinfo{person}{Koustuv Saha}, \bibinfo{person}{Eric Gregori}, \bibinfo{person}{David Joyner}, {and} \bibinfo{person}{Ashok Goel}.} \bibinfo{year}{2021}\natexlab{}.
\newblock \showarticletitle{Towards mutual theory of mind in human-ai interaction: How language reflects what students perceive about a virtual teaching assistant}. In \bibinfo{booktitle}{\emph{Proceedings of the 2021 CHI conference on human factors in computing systems}}. \bibinfo{pages}{1--14}.
\newblock


\bibitem[\protect\citeauthoryear{Wei, Shuster, Szlam, Weston, Urbanek, and Komeili}{Wei et~al\mbox{.}}{2023}]%
        {wei2023multi}
\bibfield{author}{\bibinfo{person}{Jimmy Wei}, \bibinfo{person}{Kurt Shuster}, \bibinfo{person}{Arthur Szlam}, \bibinfo{person}{Jason Weston}, \bibinfo{person}{Jack Urbanek}, {and} \bibinfo{person}{Mojtaba Komeili}.} \bibinfo{year}{2023}\natexlab{}.
\newblock \showarticletitle{Multi-party chat: Conversational agents in group settings with humans and models}.
\newblock \bibinfo{journal}{\emph{arXiv preprint arXiv:2304.13835}} (\bibinfo{year}{2023}).
\newblock


\bibitem[\protect\citeauthoryear{Wei, Wang, Schuurmans, Bosma, Xia, Chi, Le, Zhou, et~al\mbox{.}}{Wei et~al\mbox{.}}{2022}]%
        {wei2022chain}
\bibfield{author}{\bibinfo{person}{Jason Wei}, \bibinfo{person}{Xuezhi Wang}, \bibinfo{person}{Dale Schuurmans}, \bibinfo{person}{Maarten Bosma}, \bibinfo{person}{Fei Xia}, \bibinfo{person}{Ed Chi}, \bibinfo{person}{Quoc~V Le}, \bibinfo{person}{Denny Zhou}, {et~al\mbox{.}}} \bibinfo{year}{2022}\natexlab{}.
\newblock \showarticletitle{Chain-of-thought prompting elicits reasoning in large language models}.
\newblock \bibinfo{journal}{\emph{Advances in neural information processing systems}}  \bibinfo{volume}{35} (\bibinfo{year}{2022}), \bibinfo{pages}{24824--24837}.
\newblock


\bibitem[\protect\citeauthoryear{Weizenbaum}{Weizenbaum}{1966}]%
        {eliza1966}
\bibfield{author}{\bibinfo{person}{Joseph Weizenbaum}.} \bibinfo{year}{1966}\natexlab{}.
\newblock \showarticletitle{ELIZA—a Computer Program for the Study of Natural Language Communication between Man and Machine}.
\newblock \bibinfo{journal}{\emph{Commun. ACM}} \bibinfo{volume}{9}, \bibinfo{number}{1} (\bibinfo{date}{jan} \bibinfo{year}{1966}), \bibinfo{pages}{36–45}.
\newblock
\showISSN{0001-0782}
\urldef\tempurl%
\url{https://doi.org/10.1145/365153.365168}
\showDOI{\tempurl}


\bibitem[\protect\citeauthoryear{Woodruff and Aoki}{Woodruff and Aoki}{2003}]%
        {woodruff2003push}
\bibfield{author}{\bibinfo{person}{Allison Woodruff} {and} \bibinfo{person}{Paul~M Aoki}.} \bibinfo{year}{2003}\natexlab{}.
\newblock \showarticletitle{How push-to-talk makes talk less pushy}. In \bibinfo{booktitle}{\emph{Proceedings of the 2003 ACM International Conference on Supporting Group Work}}. \bibinfo{pages}{170--179}.
\newblock


\bibitem[\protect\citeauthoryear{Xu, Wang, Niu, Wu, Che, and Liu}{Xu et~al\mbox{.}}{2020}]%
        {xu2020conversational}
\bibfield{author}{\bibinfo{person}{Jun Xu}, \bibinfo{person}{Haifeng Wang}, \bibinfo{person}{Zheng-Yu Niu}, \bibinfo{person}{Hua Wu}, \bibinfo{person}{Wanxiang Che}, {and} \bibinfo{person}{Ting Liu}.} \bibinfo{year}{2020}\natexlab{}.
\newblock \showarticletitle{Conversational graph grounded policy learning for open-domain conversation generation}. In \bibinfo{booktitle}{\emph{Proceedings of the 58th annual meeting of the association for computational linguistics}}. \bibinfo{pages}{1835--1845}.
\newblock


\bibitem[\protect\citeauthoryear{Yamashita, Inoue, Guo, Mochizuki, Kawahara, and Higashinaka}{Yamashita et~al\mbox{.}}{2023}]%
        {yamashita2023realpersonachat}
\bibfield{author}{\bibinfo{person}{Sanae Yamashita}, \bibinfo{person}{Koji Inoue}, \bibinfo{person}{Ao Guo}, \bibinfo{person}{Shota Mochizuki}, \bibinfo{person}{Tatsuya Kawahara}, {and} \bibinfo{person}{Ryuichiro Higashinaka}.} \bibinfo{year}{2023}\natexlab{}.
\newblock \showarticletitle{Realpersonachat: A realistic persona chat corpus with interlocutors’ own personalities}. In \bibinfo{booktitle}{\emph{Proceedings of the 37th Pacific Asia Conference on Language, Information and Computation}}. \bibinfo{pages}{852--861}.
\newblock


\bibitem[\protect\citeauthoryear{Yao, Yu, Zhao, Shafran, Griffiths, Cao, and Narasimhan}{Yao et~al\mbox{.}}{2024}]%
        {yao2024tree}
\bibfield{author}{\bibinfo{person}{Shunyu Yao}, \bibinfo{person}{Dian Yu}, \bibinfo{person}{Jeffrey Zhao}, \bibinfo{person}{Izhak Shafran}, \bibinfo{person}{Tom Griffiths}, \bibinfo{person}{Yuan Cao}, {and} \bibinfo{person}{Karthik Narasimhan}.} \bibinfo{year}{2024}\natexlab{}.
\newblock \showarticletitle{Tree of thoughts: Deliberate problem solving with large language models}.
\newblock \bibinfo{journal}{\emph{Advances in Neural Information Processing Systems}}  \bibinfo{volume}{36} (\bibinfo{year}{2024}).
\newblock


\bibitem[\protect\citeauthoryear{Yao, Zhao, Yu, Du, Shafran, Narasimhan, and Cao}{Yao et~al\mbox{.}}{2022}]%
        {yao2022react}
\bibfield{author}{\bibinfo{person}{Shunyu Yao}, \bibinfo{person}{Jeffrey Zhao}, \bibinfo{person}{Dian Yu}, \bibinfo{person}{Nan Du}, \bibinfo{person}{Izhak Shafran}, \bibinfo{person}{Karthik Narasimhan}, {and} \bibinfo{person}{Yuan Cao}.} \bibinfo{year}{2022}\natexlab{}.
\newblock \showarticletitle{React: Synergizing reasoning and acting in language models}.
\newblock \bibinfo{journal}{\emph{arXiv preprint arXiv:2210.03629}} (\bibinfo{year}{2022}).
\newblock


\bibitem[\protect\citeauthoryear{Zhang}{Zhang}{2018}]%
        {zhang2018personalizing}
\bibfield{author}{\bibinfo{person}{Saizheng Zhang}.} \bibinfo{year}{2018}\natexlab{}.
\newblock \showarticletitle{Personalizing dialogue agents: I have a dog, do you have pets too}.
\newblock \bibinfo{journal}{\emph{arXiv preprint arXiv:1801.07243}} (\bibinfo{year}{2018}).
\newblock


\end{thebibliography}


\section{Appendix}

\subsection{System}
\label{apdx:playground-settings}
\begin{figure*}[ht]
\centering
\includegraphics[width=\linewidth]{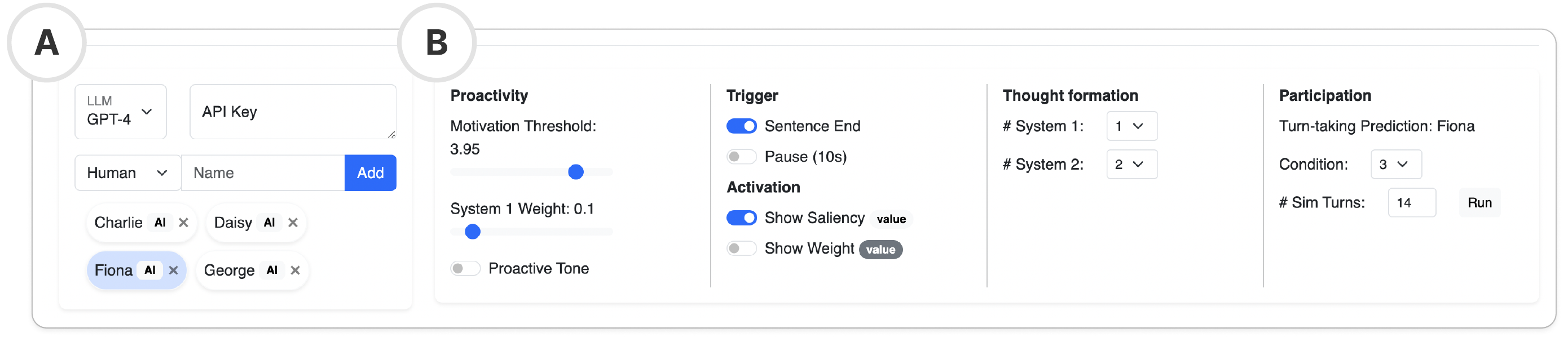}
\caption{Settings page for Inner Thoughts Playground web app. Users can change the LLM version, add/delete conversation participants, adjust proactivity level, trigger method, thought formation quantity and participation strategies.}
\label{fig:playground_settings}
\Description{Settings page for Inner Thoughts Playground web app. Users can change the LLM version, add/delete conversation participants, adjust proactivity level, trigger method, thought formation quantity and participation strategies.}
\end{figure*}

~\autoref{fig:playground_settings} shows the Inner Thoughts playground settings panel.

\subsection{User Evaluation}
\label{apdx:user-eval}

\subsubsection{Proactivity Settings}

\begin{enumerate}
    \item \textit{Non-stop chatter}: This AI engaged continuously in the conversation, even when it had little relevant input to offer.
    \begin{itemize}
        \item \texttt{system1Prob} = 0.7
        \item \texttt{imThreshold} = 4.49
        \item \texttt{interruptThreshold} = 4.8
        \item \texttt{proactiveTone} = false
    \end{itemize}
    
    \item \textit{Active contributor}: This AI participated actively, contributing when appropriate but without dominating the conversation.
    \begin{itemize}
        \item More proactive AI
        \item \texttt{system1Prob} = 0.2
        \item \texttt{imThreshold} = 3.59
        \item \texttt{interruptThreshold} = 4.8
        \item \texttt{proactiveTone} = true
    \end{itemize}
    
    \item \textit{Selective participant}: This AI contributed only when highly interested in the topic, remaining silent during other parts of the conversation.
    \begin{itemize}
        \item Less proactive AI
        \item \texttt{system1Prob} = 0
        \item \texttt{imThreshold} = 4.09
        \item \texttt{interruptThreshold} = 5
        \item \texttt{proactiveTone} = false
    \end{itemize}
\end{enumerate}

\subsubsection{Metrics}
\begin{table*}[]
\resizebox{\textwidth}{!}{\begin{tabular}{@{}ll@{}}
\toprule
\textbf{Metric}           & \textbf{Statement}                                                 \\ \midrule
Anthropomorphism          & I felt the chatbot is humanlike                                    \\
Likeability               & I felt pleasant to chat with the chatbot                           \\
Initiative                & I felt the chatbot is able to take the initiative in conversations \\
Perceived social presence & I was often aware of the chatbot in our conversation               \\
Perceived engagement                   & I could feel that chatbot is engaging well in the conversation                                \\
Perceived listener        & I felt chatbot was actively listening and I was heard              \\
Contribution                           & The chatbot made valuable contributions that enhanced the overall quality of the conversation \\
Appropriateness of the response timing & I felt that the chatbot can join in conversation at appropriate moments                       \\
Future usage                           & I’d like to have this chatbot when have similar conversation in the future                    \\
Extroversion              & I felt like the chatbot has an extroverted personality             \\ \bottomrule
\end{tabular}}
\caption{Metrics used in our user study to measure the quality of AI simulated conversations. Each statement is rated on a Likert-scale from 1 -- Strongly Disagree to 7 -- Strongly Agree.}
\label{tab:metrics-user}
\end{table*}

\autoref{tab:metrics-user} shows the metrics and their definitions that we used in our user study.

\end{document}